\documentclass{article}

\usepackage{arxiv}

\usepackage[utf8]{inputenc} % allow utf-8 input
\usepackage[T1]{fontenc}    % use 8-bit T1 fonts
\usepackage{hyperref}       % hyperlinks
\usepackage{url}            % simple URL typesetting
\usepackage{booktabs}       % professional-quality tables
\usepackage{amsfonts}       % blackboard math symbols
\usepackage{nicefrac}       % compact symbols for 1/2, etc.
\usepackage{microtype}      % microtypography
\usepackage{lipsum}		% Can be removed after putting your text content
\usepackage{graphicx}
\usepackage{natbib}
\usepackage{doi}
\usepackage{float}

\usepackage{algorithm}
\usepackage{algpseudocode}
\usepackage{amsmath}
\usepackage{amssymb}
\usepackage{bbm}
\usepackage{caption}
\usepackage{subcaption}

\title{Estimating the Instantaneous Reproduction Number With Imperfect Data: A Method to Account for Case-Reporting Variation and Serial Interval Uncertainty}

\author{Gary Hettinger \\
	Division of Biostatistics \\
	University of Pennsylvania\\
	Philadelphia, PA, U.S.A. \\
	\texttt{ghetting@pennmedicine.upenn.edu} \\
	\And
	David Rubin \\
	Perelman School of Medicine\\
	University of Pennsylvania\\
	Philadelphia, PA, U.S.A. \\
        \And
	Jing Huang \\
	Division of Biostatistics\\
	University of Pennsylvania\\
	Philadelphia, PA, U.S.A \\
}

\renewcommand{\shorttitle}{Estimating the Reproduction Number with Imperfect Data}

\hypersetup{
pdftitle={Estimating the Instantaneous Reproduction Number With Imperfect Data: A Method to Account for Case-Reporting Variation and Serial Interval Uncertainty},
pdfauthor={Gary~Hettinger, David~Rubin, Jing~Huang},
pdfkeywords={Bayesian, COVID-19, Infectious Disease},
}

\begin{document}
\maketitle

\begin{abstract}
During an infectious disease outbreak, public health decision-makers require real-time monitoring of disease transmission to respond quickly and intelligently. In these settings, a key measure of transmission is the instantaneous time-varying reproduction number, $R_t$. Estimation of this number using a Time-Since-Infection model relies on case-notification data and the distribution of the serial interval on the target population. However, in practice, case-notification data may contain measurement error due to variation in case reporting while available serial interval estimates may come from studies on non-representative populations. 

We propose a new data-driven method that accounts for particular forms of case-reporting measurement error and can incorporate multiple partially representative serial interval estimates into the transmission estimation process. In addition, we provide practical tools for automatically identifying measurement error patterns and determining when measurement error may not be adequately accounted for. We illustrate the potential bias undertaken by methods that ignore these practical concerns through a variety of simulated outbreaks. We then demonstrate the use of our method on data from the COVID-19 pandemic to estimate transmission and explore the relationships between social distancing, temperature, and transmission. 
\end{abstract}

% keywords can be removed

\section{Introduction}
\label{sec:intro}

During infectious disease outbreaks, decision-makers require timely, accurate monitoring of disease transmission to inform public health responses. An important quantity for infectious disease surveillance is the reproduction number, $R$, which is defined as the average number of secondary cases generated by an infected individual over his or her infectious period. As demonstrated during the recent COVID-19 pandemic, the quantity has many uses including to monitor and project the spread of disease in a particular region (\cite{WangSong2020, Chen2020, Wu2020}), parameterize models that evaluate intervention efficacy (\cite{Pan2020, JorgeBrazil2021, Zhang2021, PengIntervention2020}), and provide insights into factors associated with transmission (\cite{Rubin2020, WangTeunis2020}). When real-time monitoring is a priority, \cite{Fraser2007} and \cite{Gostic2020} recommend estimating the time-varying instantaneous reproduction number, $R_t$, which represents the average number of secondary cases that an individual infected at time $t$ could expect to infect should current conditions persist.

Numerous statistical methods have been developed to estimate the reproduction number using the Time-Since-Infection (TSI) epidemiological model (\cite{Kermack1927}), with many summarized in \cite{White2021}. When estimating time-varying reproduction numbers, methodologists have largely favored Bayesian approaches due to complex likelihoods induced in the time-varying setting (\cite{CauchemezRealTime2006, Cori2013, ThompsonCori2019, QuickLin2021}). In particular, \cite{Cori2013}'s development of an estimator for $R_t$ has been widely used in practice, including during the COVID-19 pandemic.

Despite the success of widely-used methodology, TSI models inherently rely on case-notification data, which is known to contain measurement error due to variation in case-reporting (\cite{Gostic2020, RiconBeckerWeekly2020, Bergman2020OscillationsFactors}). High reporting variation often renders $R_t$ estimates too volatile for practical use. In these settings, practitioners commonly aggregate case counts or transmission estimates over a sliding window of pre-determined length (\cite{Gostic2020}). However, these smoothing techniques are not targeted to reporting patterns and therefore may smooth over relevant signals in the data. Previous methods have addressed limitations in case ascertainment when additional seroprevalance (\cite{QuickLin2021}) or hospitalization data (\cite{WhiteInfluenza2009}) are available and accurate. \cite{Shi2022RobustSARS-CoV-2} develop a flexible approach that allows for uncertainty in transmission estimates not accounted for in a predictive time-series model, which may result in part from case reporting patterns, but do not directly target nuisance reporting signals in the observed data.

In addition to case-notification data, TSI models require an estimate for the distribution of the serial interval, which is defined as the time between case definitions of infector-infected pairs. Particularly early in an outbreak, serial interval estimates may come from small studies on a particular subpopulation, often close to the outbreak source. However, prior work has found significant heterogeneity in serial intervals across many factors including region (\cite{RaiSerialIntervals2021}, \cite{AleneSerialInterval2021}) and age (\cite{ McAloonIrelandSI2021}). Therefore, a single serial interval study is likely representative of only a fraction of the target population. Researchers have tried to address this limitation by allowing for additional uncertainty of this parameter in their models. \cite{Cori2013} allowed for serial interval uncertainty with a grid-sampling approach over a user-specified range of serial interval means and variances. \cite{ThompsonCori2019} improved upon this approach with a two-step process that first estimates the posterior distribution of the serial interval using known pairs of index and secondary cases and then uses this posterior distribution to estimate $R_t$, but relies on representative and accurate contact tracing data.

Here, we propose a unified framework to robustly estimate $R_t$ in the presence of case reporting measurement error and non-representative serial interval studies without additional contact tracing, seroprevalence, or hospitalization data. Our Bayesian framework joins together a model of case-reporting patterns to address measurement error, a mixture model to incorporate multiple serial interval distributions, and an autoregressive transmission model that dually serves to increase statistical power and identify factors associated with transmission. To assist practitioners, we additionally provide an algorithm to identify weekly reporting patterns and an objective test to determine when pre-analysis case-smoothing may be favorable to a measurement error model.

The remainder of the paper is organized as follows: In Section~\ref{sec:methods}, we first describe the TSI model before introducing our proposed model. In Section~\ref{sec:simulations}, we present numerical studies across a variety of simulated outbreaks to empirically demonstrate the added capabilities of our proposed approach. We then apply our methodology to study the United States COVID-19 outbreak in 2020 to estimate transmission and understand transmission risk factors in Section~\ref{sec:application}. Finally, we summarize our findings and discuss further methodological considerations in Section~\ref{sec:discussion}.

\section{Methods}
\label{sec:methods}

\subsection{Time-Since-Infection Model}

\label{sec:methods:tsi}

We first briefly review the foundations of TSI model estimators for $R_t$ as described in \cite{Fraser2007}. Let $I_t$ denote the number of incident infections generated on day $t$ and $\beta_{t,s}$ be the infectiousness on day $t$ for an individual that was infected $s$ days ago. The possible generators of $I_t$ are the previously infected individuals: $I_{t-1},...,I_0$, where an individual infected on day $t-s$ generates, on average, $\beta_{t,s}$ new infections on day $t$. We can then specify $I_t$ as a function of $\beta_{t,s}$ and $I_{t-1},...,I_0$ through the so-called renewal equation:

\begin{equation}
    \label{eqn:renewal}
    I_t = \sum\limits_{s=1}^{S^*} \beta_{t,s} I_{t-s}
\end{equation}
where $S^*$ denotes the number of days at which infectivity is considered no longer viable. Assuming that the generation interval distribution, $w_s$, defined as the proportion of total infections generated by an infected individual that are generated on day $s$ since infection, is independent of calendar time, $t$, we can decompose $\beta_{t,s}$ as:

\begin{equation}
    \label{eqn:infectiousness_separation}
    \beta_{t,s} = R_t w_s
\end{equation}
where $R_t$ is our instantaneous reproduction number. Combining (\ref{eqn:renewal}) and (\ref{eqn:infectiousness_separation}), we can solve for $R_t$ as:

\begin{equation}
    \label{eqn:rt_estimator}
    R_t = \frac{I_t}{\sum\limits_{s} w_s I_{t-s}}
\end{equation}
In practice, we do not observe the true generation interval or infections, leading to probabilistic methods that allow for uncertainty in this relationship.

\subsection{A Bayesian Model in the Presence of Case Reporting Measurement Error and Serial Interval Distribution Uncertainty}
\label{sec:methods:model}

Consider observing a time-series of reported cases on day of outbreak $t$, $I_t$, which are a proxy of the true latent incident cases, $I^*_t$, where $t=1, \dots, n$. The observed and latent cases differ due to reporting patterns that are consistent within reporting periods, $\tau(t)$ (e.g., day-of-week). We observe a vector of time-varying covariates, $\mathbf{X_t}$, associated with the latent instantaneous reproduction number on day $t$, $R_t$. Finally, we observe $K$ external serial interval distribution estimates, $\mathbf{w}^{(k)}$ ($k=1,\dots, K$), included to represent at least part of the target population. 

Then, we specify our model over the joint data distribution hierarchically as:
\begin{align}
    & I_t |  I^*_t, \boldsymbol{\eta} \sim Poisson(\theta_{\tau(t)} I^*_t) \label{eqn:me_model}\\
    & I^*_t | I^*_{1:(t-1)}, R_t, \mathbf{w^*} \sim Poisson(R_t \sum\limits_{s=1}^{S^*} w^*_s I^*_{t-s}) \label{eqn:ext_tsi_model}\\
    & R_t |R_{1:(t-1)}, \mathbf{X}, \boldsymbol{\eta} \sim GARMA(\mathbf{X_t}, AR=m, MA=q, \text{family}=Gamma)  \label{eqn:trans_model} \\
    & \hspace{2.5mm} \boldsymbol{\eta}  | G, \boldsymbol{\gamma} \sim G(\gamma) \label{eqn:priors}
\end{align}

In the first layer of our model, we specify that the observed reported cases, $I_t$, are centered around the scaled latent true incident cases, where the scaling weights, $\theta_{\tau(t)}$, are specific to the reporting period of day $t$, $\tau(t)$. 

We then link incident cases to $R_t$ by extending the TSI model to incorporate a mixture of serial interval distribution estimates, 
$w^*_s = \sum\limits_{k} \lambda_k w^{(k)}_s$. Here, $\lambda_1,...,\lambda_K$ are unknown mixture weights that allow for uncertainty in the representativeness of available estimates.

We propose modeling $R_t$ according to the generalized autoregressive moving average model (GARMA) derived in \cite{BenjaminGarma2003} for a \textit{Gamma} distribution with autoregressive order $m$ and moving average order $q$. Setting $q=0$, this model specifies:
\begin{equation*}
    E[R_t] = exp\{\boldsymbol{\beta}^\top \mathbf{X_t} + \sum_{i=1}^m \phi_i[log(R_{t-i}) - \boldsymbol{\beta}^\top \mathbf{X_{t-i}})]\}
\end{equation*}
where $\boldsymbol{\beta}$ and $\boldsymbol{\phi}$ denote the unknown covariate-effect and autoregressive parameters, respectively. By incorporating an autoregressive model for $R_t$, we can improve the stability and power of our estimates in a statistically-principled manner, while also enabling the exploration of associations between predictors and transmission (\cite{Shi2022RobustSARS-CoV-2}). 

We define $\boldsymbol{\eta} = (\boldsymbol{\theta}, \boldsymbol{\lambda}, \boldsymbol{\beta}, \boldsymbol{\phi}, \sigma_R)$ for compactness, where these are drawn from a distribution $G$ composed of generally noninformative independent priors according to hyper-parameters $\gamma$.

\subsection{Determining the Case Reporting Measurement Error Model}
\label{sec:methods:me_model}
Realistically, practitioners may have to identify the mapping $\tau(t)$ in (\ref{eqn:me_model}) that partitions day-of-outbreak into reporting periods themselves. An ideal mapping captures all variation in cases due to reporting patterns and none of the variation in cases due to important transmission signals. Here, we propose developing a mapping based on day-of-week patterns, which are highly associated with reported case counts and appear likely to persist in future outbreaks (\cite{RiconBeckerWeekly2020, Bergman2020OscillationsFactors}). While relevant transmission factors like increased socialization on weekends may explain some of the day-of-week patterns, we can incorporate social distancing metrics into (\ref{eqn:trans_model}) to capture the relevant components of these signals. 

When model parsimony is important (e.g., shorter length studies), we propose the following algorithm to automatically partition the days of the week into periods of relatively homogeneous reporting patterns:

\begin{algorithm}
\label{alg:reporting_pattern}
  \caption{Day-of-Week (DOW) Reporting Pattern Detection Algorithm}\label{alg:dow_alg}
  \begin{algorithmic}[1]
    \Procedure{ReportingPatternDetection}{$\mathbf{I}$}
        \State $I^*_t \gets \sum\limits_{i=-2}^{2} I_{t-i}$\Comment{Rolling average as proxy for $I^*_t$}
        \State $V_t \gets I_t / I^*_t$\Comment{Proxy for relative reporting variation}
        \State $\boldsymbol{\nu^{(l)}} \gets \{ V_t : DOW(t) = l \}$\Comment{Vector of variation for day-of-week $l$ indexed by week}
        \For{$K=1:7$}
            \State $\tau^{(K)}(l) \gets \text{kMeansCluster}(\begin{bmatrix}
            \boldsymbol{\nu^{(Su)}}^\top\\
            \vdots \\
            \boldsymbol{\nu^{(Sa)}}^\top
            \end{bmatrix}, k=K)$\Comment{Cluster the days of week by reporting variation}
            \State $\text{model}^{(K)} \gets linreg(\nu^{(l)}_j \sim \theta^{(K)}_1 C_1(l) + \dots + \theta^{(K)}_K C_K(l))$\Comment{$C_k(l) = \mathbbm{1}\{\tau^{(K)}(l) =  k\}$} 
        \EndFor
        \State $K^* \gets \underset{K}{\operatorname{argmax}} (AIC(\text{model}^{(k)})$\Comment{Select clustering that minimizes AIC}
        \State $\tau(t) \gets \tau^{(K^*)}(DOW(t))$\Comment{Choose optimal mapping}
        \State $\theta^{(prior)}_{\tau(t)} = \theta^{(K^*)}_{\tau(t)}$\Comment{Set prior weights as coefficients from regression model}
    \EndProcedure
  \end{algorithmic}
\end{algorithm}

Intuitively, Algorithm~\ref{alg:dow_alg} first identifies general temporal trends in cases (Line 2), which allows us to then separate out the day-of-week patterns (Line 3) using standard time-series decompositions (\cite{Forecastingtextbook}). We then cluster the days of the week (Line 6) according to a vector of their relative reporting variation (Line 4). With each size-$K$ clustering, we regress the reporting variations on cluster-specific intercepts to test how well the clustering explains the reporting variation (Line 7). Finally, we select a parsimonious clustering of the days of week according to the Akaike Information Criterion (AIC) (Line 9), which gives us our mapping $\tau(t)$ (Line 10). A visualization of this method using COVID-19 data from New York is provided in Figure A1. 

The estimated cluster-specific coefficients (Line 11) provide reasonable informative priors for the reporting weights, $\theta_{\tau(t)}$, if desired. A perhaps more important consideration for the priors on these weights is the choice in prior distribution, which may enforce certain constraints as discussed further in Section A1. In practice, we constrain the mean of the weights to be equal to one through a Dirichlet prior. 

In certain settings, the day-of-week measurement error model may not sufficiently separate reporting patterns from transmission signals, like when reporting patterns change over time and are correlated with $R_t$. In lieu of an alternative sufficient model, it may be necessary for practical utility to instead reduce measurement error with a more coarse smoothing technique, like by taking rolling averages of case data prior to analysis (\cite{Gostic2020}). Whereas the method of \cite{Cori2013} uses window-based techniques to smooth $R_t$ estimates, we choose to smooth reported cases to more directly target case reporting measurement error. Since coarse smoothing approaches may smooth over relevant transmission signals, we suggest pre-analysis case-smoothing only when it provides a substantial improvement to the volatility of estimates, $\hat{R}_t$. If $\hat{R}_t$ using the day-of-week measurement error model is substantially more volatile than when using case-smoothing, then it is likely that $\hat{R}_t$ using the day-of-week measurement error model is too volatile to be useful. We measure volatility via lag-one auto-correlation, which estimates the correlation between $R_t$ and $R_{t-1}$ across $t$, and ranges between -1 (high volatility) to +1 (low volatility). To make the decision between measurement error approaches objectively in our work, we derive $95\%$ confidence intervals for the lag-one auto-correlation of the $\hat{R}_t$ time-series from the measurement error model approach using corrected Pearson coefficient variances (\cite{Afyouni2019}) and only select the case-smoothing method when the lag-one auto-correlation of $\hat{R}_t$ time-series from the case-smoothing approach falls above the upper confidence bound. 

\section{Simulation Studies}
\label{sec:simulations}

\subsection{Design}
\label{sec:simulations:design}

We designed our simulations to mimic the noted imperfections in available real-world data gathered from disease outbreaks. We simulated outbreaks across a variety of realistic \textit{trend patterns} (TPs), which specify how transmission changes over the outbreak period. These temporal patterns corresponded to three intuitive phases: an \textit{unrestricted} growth phase representing a period of high transmission (e.g., minimal public response), a \textit{lockdown} phase where transmission decreases (e.g., high public response), and an \textit{equilibrium} phase where balance occurs between the virus and public responses. 

TP1 outbreaks followed an \textit{unrestricted-lockdown-equilibrium} paradigm, whereas TP2 outbreaks followed an \textit{unrestricted-lockdown-unrestricted-lockdown-equilibrium} transmission paradigm. In TP3 outbreaks, transmission transitioned smoothly from \textit{unrestricted} to \textit{equilibrium}. These transmission patterns were generated by simulating:
\begin{equation*}
	X_t \sim Unif(-0.05, 0.05) + \begin{cases} \mu_{phase} & \text{Context 1,2} \\ \frac{-0.7}{1+exp(-0.25(t-15))}-0.1 & \text{Context 3} \end{cases}
\end{equation*}
where $\mu_{unrestricted}=-0.2$, $\mu_{lockdown}=-0.85$, $\mu_{equilibrium}= -0.54$. $X_t$ is designed to align with the \textit{daily visitation difference} social distancing measure described further in Section~\ref{sec:application}. 

After specifying $\tau(t)$, $\mathbf{w^{(1:K)}}$, and $\boldsymbol{\eta}$ parameters, we then sequentially simulated $R_t$ according to (\ref{eqn:trans_model}), $I^*_t$ according to (\ref{eqn:ext_tsi_model}), and $I_t$ according to (\ref{eqn:me_model}). Here, $\tau(t)$ generally mapped day-of-outbreak to period-of-week clusters - $\{Sa,Su\}$, $\{M, Tu\}$, $\{W,Th,F\}$. Specified parameters for $\mathbf{w^{(1:K)}}$ and $\boldsymbol{\eta}$ can be found in Section A2. Examples of simulated $R_t$ time-series by trend pattern are visualized in Figure A2.

In addition to outbreak TPs, we also differed simulations by \textit{data scenarios} (DSs), which represent different nuances in the observed data and are summarized in Table~\ref{tab:scenarios}. In DS0, we generated data with no variation in case reporting ($I_t=I^*_t$), a single (correctly-specified) serial interval distribution ($\mathbf{w^{(1)}} = \mathbf{w^*}$, $K=1$), and provided our model an uninformative observed covariate ($X^{(obs)}_t \sim Unif(-0.1, 0.1) \neq X_t$). Therefore, our proposed model only extends the method of \cite{Cori2013} by including a misspecified model for (\ref{eqn:trans_model}). In each sub-scenario of DS1, we added a single nuance to the observed data – (A) multiple serial interval interval estimates are available ($\mathbf{w^*} = \lambda_1 \mathbf{w^{(1)}} + \lambda_2 \mathbf{w^{(2)}} + \lambda_3 \mathbf{w^{(3)}}$), (B) the true data-generating mechanism covariate is observed ($X_t = X^{(obs)}_t$), and (C) observed reported cases contain measurement error ($\boldsymbol{\theta} = (0.5, 1.5, 1.0)$).

DS2 combines all three nuances of DS1, while simulating from the parametric distributions specified in our model (DS2A) and alternative distributions (DS2B). In the latter, we violated the parametric assumptions in (\ref{eqn:me_model}), (\ref{eqn:ext_tsi_model}), and (\ref{eqn:trans_model}) by generating $I_t$, $I^*_t$, and $R_t$ from Log-Normal (rounded), Negative Binomial, and Log-Normal distributions, respectively.

\begin{table}
    \caption{Simulation scenario nuances.}
    \label{tab:scenarios}
    \centering
    \begin{tabular}{lllll}
    \hline
     Data & Reporting & Serial & Predictors & Parametric  \\
    Scenario & Variation & Interval & Specified & Specification \\
    \hline
    0  & Null & Given & No & Correct\\
	1A  & Null & Uncertain & No & Correct\\
	1B  & Null & Given & Yes & Correct\\
	1C  & Correctly Specified & Given & No & Correct\\
	2A  & Correctly Specified & Uncertain & Yes & Correct\\
	2B  & Correctly Specified & Uncertain & Yes & Misspecified\\
	3A,B & Misspecified & Uncertain & Yes & Correct\\
	3C & Misspecified & Given & Yes & Correct\\
    \hline
    \end{tabular}
\end{table}

 In each sub-scenario of DS3, we generated observed case data to violate a different assumption of our measurement error model – (A) reporting patterns stop halfway through the outbreak ($\boldsymbol{\theta} = (0.5, 1.5, 1.0) \rightarrow (1,1,1)$), violating our time-invariant assumption on $\boldsymbol{\theta}$ and (B) reporting weights do not have mean 1 ($\boldsymbol{\theta} = (0.5, 1.25, 0.8)$), violating the Dirichlet prior we impose on $\boldsymbol{\theta}$. In DS3C, reporting patterns cannot be separated from transmission signals as case reporting alternates between low reporting on odd days and high reporting on the following day (intuitively to catch up), does not have mean 1 weights, and is correlated with $R_t$. We implemented this by setting $\tau(t) = \text{parity}(t)$, $\boldsymbol{\theta} = (0.5, 1.23)$ in the unrestricted phase, $\boldsymbol{\theta} = (0.5, 2)$ in the lockdown phase, and $\boldsymbol{\theta} = (0.5, 1.5)$ in the equilibrium phase.

\subsection{Implementation}
\label{sec:simulations:implementation}

For DS0-DS3B, we generated short simulations (50 days) for all trend patterns and long simulations (100 days) for TP2. For DS3C, we only generated short simulations for TP1. We conducted 1200 simulations for each scenario-trend pattern pair and discarded simulations with low case counts ($I_t < 10$) for more than $20\%$ of days as these generally indicated outbreaks that did not persist for the entire study length. On average, a scenario-context pair resulted in 989 and 804 kept simulations for short and long runs, respectively. 

We evaluated the relative performance of $\hat{R}_t$ and 95\% credible intervals from our proposed method and the method of \cite{Cori2013} (Baseline) using Mean Squared Error (MSE), Mean Raw Bias (Bias), and coverage probability. For MSE and Bias, we first scaled the difference between estimated and true values by the true parameter value. We did not consider the first 7 days of estimates to restrict model comparisons to estimates with sufficient case data. 

We implemented the proposed model with \textit{JAGS} (\cite{JAGS}) to develop a Markov Chain Monte Carlo (MCMC) sampler with 8 chains, 2000 burn-in iterations and 4000 total iterations per chain, and a thinning parameter of 2. We used the posterior mean for point estimates of parameters and estimated $95\%$ credible intervals using the $2.5$ and $97.5$ percentiles of the posterior samples. We implemented the baseline method with the \textit{EpiEstim} R package (\cite{Cori2021EpiEstimCurves}) and input the mean of the available serial interval estimates in data scenarios where the correct serial interval was not considered given (DS1A, DS2A-DS3B).

\subsection{Results}
\label{sec:simulations:results}

We first compared the performance of baseline method implementations according to the window size, or bandwidth, parameter ($W=1,3,5,7$) and specification of serial interval uncertainty (U) to select an appropriate comparison to our method. We centered windows at $t$ (i.e., window=$[t- \frac{W}{2} , t + \frac{W}{2}]$) to estimate $R_t$, since these resulted in estimates most aligned with the true $R_t$.

Results from the $W=1 (U)$, $W=1$, and $W=7$ models are summarized in Table A1. The choice in $W$, which controls the smoothness of $\hat{R}_t$ as visualized in Figure A3(a), drastically altered model performance depending on which assumptions hold. In DS0-DS1B, models with $W=1$ performed best in terms of MSE, bias, and coverage probability since they do not smooth over the signals in case data, which are largely relevant in these simple scenarios. However, smoothing with $W=7$ resulted in comparable (DS2A-DS3B) and even improved (DS3C) estimates in more complex settings when reporting patterns induced large amounts of noise in case data. The approach of the baseline method to address serial interval uncertainty (U) improved coverage probability in our simulations but drastically increased MSE and Bias of $\hat{R}_t$. The low point estimation accuracy for this approach is likely influenced by the disconcordance between the true serial interval distribution form (max 4 days, non-parametrically distributed) and the assumed form of the method (max 20 days, Gamma distributed). Therefore, we focused on the baseline model with $W=1$ and serial interval distribution uncertainty ignored, since this window size is most similar in spirit to our attempt to capture granular signals in $R_t$ and this implementation had the best performance for a majority of scenarios.\\

\underline{Comparison between Proposed and Baseline Methods with Specified Reporting Periods}

\begin{table}
	\caption{Instantaneous reproduction number estimation comparison between baseline and proposed models for simulations with well-specified reporting periods. Relative change is from baseline method performance.}
	\centering
	\begin{tabular}{l|ll|ll|ll}
		\hline
		Data & \multicolumn{2}{c}{MSE (x$10^{-2}$)} & \multicolumn{2}{c}{Bias ($\%$)} & \multicolumn{2}{c}{Coverage Probability ($\%$)} \\
		\cline{2-3} \cline{4-5} \cline{6-7}
		Scenario & Propose & Relative Change ($\%$) & Proposed & Relative Change ($\%$)  & Proposed & Baseline \\
		\hline
        0 & 1.29 & -10.4 & 1.88 & -6.5 & 94.97 & 95.13 \\
        1A & 1.48 & -43.5 & 2.04 & +5.7  & 95.41 & 70.65 \\
        1B & 1.36 & -34.0 & 2.42 & -24.4 & 94.88 & 95.07 \\
        1C & 3.78 & -91.0 & 3.57 & -65.5  & 90.91 & 16.86\\
        2A & 4.08 & -89.6 & 4.88 & -51.9  & 91.78 & 19.80\\
        2B & 11.99 & -77.3 & 1.63 & -80.4 & 59.93 & 16.25 \\
        3A & 10.43 & -57.3  & 6.77 & +19.0 & 70.25 & 37.85 \\
        3B & 3.79 & -87.8 & 4.97 & -42.3 & 92.92 & 22.61 \\
        3C & 212.89 & -21.8 & 69.43 & -11.3  & 0.65 & 0.72 \\
		\hline
	\end{tabular}
	\label{tab:rcomp}
\end{table}

Performance metrics for $\hat{R}_t$ comparing the proposed method with a pre-specified $\tau(t)$ to the selected baseline method are summarized in Table~\ref{tab:rcomp}. The proposed model showed almost uniformly superior MSE, bias, and coverage probability than the baseline method in these scenarios. 

In DS0-DS1B, our proposed model demonstrated improvement over the baseline method in terms of MSE (10.4-43.5\% reduction) and bias (6.5-24.4\% reduction), except for bias in DS1A (5.7\% higher), provided that we correctly ignore case reporting measurement error. Interestingly, we saw slight improvement from the baseline method in DS0 when our autoregressive model was over-specified, which speaks to the predictive power of autoregressive models in general.  
Coverage probabilities were near the nominal rate for both methods in these scenarios except for the baseline method in DS1A (70.7\% coverage), demonstrating the improvement in uncertainty quantification using our method when the true serial interval is a mixture of available serial interval estimates. The improvement in relative performance of the proposed method between DS0 and DS1B (10.4 to 34.0\% lower MSE, 6.5 to 24.4\% lower bias) suggests a benefit to modeling $R_t$ with informative covariates in the estimation process. 

When we introduced measurement error in case reporting in DS1C, we saw clear benefits of correctly modeling these patterns (91.0\% reduction in MSE, 65.5\% reduction in bias, and 74.05\% more coverage probability). These results are comparable to those in DS2A, implying a high relative importance of the measurement error model in our simulations. Figure A3(b) provides an example fit for both the proposed and baseline methods from DS2A. When we violated the parametric assumptions of our model in  DS2B, we saw diminished performance in MSE and coverage probabilities, which we attribute largely to the inability of our specified Poisson distributions to handle over-dispersion induced by the true distributions. 

In DS3A-DS3C, we explored robustness to misspecifications of the case reporting measurement error model. When we changed case reporting patterns over time in  DS3A, we saw lessened benefits (57.3\% lower MSE, 19.0\% higher bias, 32.40\% more coverage probability) since we specified a single $\boldsymbol{\theta}$ for the entire outbreak period in our model. Meanwhile, the assumption of no case reporting measurement error in the baseline method is satisfied in the second half of these outbreaks. The similar results between DS2A and DS3B suggest that imposing a constraint on the mean of $\boldsymbol{\theta}$ does not substantially reduce identification of $R_t$, even when violated. We note that this observation likely results from the partial identifiability of $\boldsymbol{\theta}$, in that only the relative reporting weights, rather than the scale of $\boldsymbol{\theta}$, affect identification of $R_t$, which we discuss further in Section A1. In  DS3C, severe misspecification of the case reporting patterns induced substantial MSE and Bias for both methods. While this scenario is likely more extreme than would be seen in practice, it suggests the need for window-based smoothing techniques when the case reporting pattern is not well identified. Figure A3(c) provides a visual example of this scenario.\\

\underline{Performance of Methods to Determine the Case Reporting Measurement Error Model}

\begin{table}
	\caption{Performance of automated measurement error model identification. Clustering accuracy is measured using a Weighted F-Measure (WFM) and aggregated by short and long runs separately. Percent of simulations with pre-analysis case smoothing and performance metrics for $R_t$ estimates using both case reporting measurement model identification and selection algorithms are averaged over short and long runs together.}
	\centering
	\begin{tabular}{l|ll|l|lll}
		\hline
		 & \multicolumn{2}{c}{Clustering Accuracy} & \multicolumn{1}{c}{Case-Smoothing} & \multicolumn{3}{c}{$R_t$ Accuracy} \\
		\cline{2-3} \cline{4-4} \cline{5-7}
		 Data & WFM & WFM  & Smoothing & MSE  & Bias & Cov. Prob.  \\
            Scenario & (Short) & (Long) & Selected ($\%$)  & ($x10^{-2})$ & ($\%$) & ($\%$) \\
		\hline
		0 & 85.8 & 84.5 & 0.2 & 6.17 & 7.72 & 89.40\\
            1A & 79.6 & 81.2 & 2.6 & 6.67 & 8.74 & 89.50\\
            1B & 85.2 & 84.0 & 0.0 & 4.14 & 7.61 & 91.34\\
            1C & 57.8 & 77.0 & 0.1 & 11.23 & 8.34 & 69.14\\
            2A & 58.7 & 77.0 & 0.2 & 10.48 & 10.60 & 72.04\\
            2B & 66.0 &  83.0 & 3.3 & 14.19 & 3.10 & 57.02\\
            3A & -- & -- & 1.0 & 17.21 & 13.40 & 56.55\\
            3B & 50.5 & 73.9 & 0.2 & 9.91 & 10.91 & 71.19\\
            3C & -- & -- & 100.0 & 18.51 & 8.78 & 23.49\\
		\hline
	\end{tabular}

        {\footnotesize -- Reporting patterns vary over time so performance metrics are not calculated}
	\label{tab:me_model_ident}
\end{table}

We also evaluated the performance of Algorithm~\ref{alg:dow_alg} in identifying the correct case reporting periods and our proposed case-smoothing decision tool in selecting an appropriate approach to address measurement error in case reporting. 

To assess the performance of Algorithm~\ref{alg:dow_alg}, we computed a \textit{weighted f-measure} (WFM) metric, which is a clustering metric ranging between 0 and 1 that looks at all pairwise combinations of the days of the week and penalizes pairs that are incorrectly clustered together (restrictive error) twice as much as those incorrectly mapped separately (parsimony error). To assess the performance of our case-smoothing decision tool, we looked at the percentage of simulations that the tool selects case-smoothing in favor of the measurement error model in settings where we prefer the former (DS3C) and the latter (DS0-DS3B). Finally, we assessed the joint performance of our model, algorithm, and decision tool in estimating $R_t$ to understand realistic method performance in practical settings. These metrics are summarized by data scenario in Table~\ref{tab:me_model_ident}.

Example WFM scores are provided in Table A2 for context. In DS0-DS1B when all days of the week should be mapped to the same cluster, the average WFM scores are between 79.6-85.8 for both short and long runs, suggesting 1-2 parsimony mistakes on average. In DS1C-DS2B and DS3B, the average WFM scores are between 50.5-66.0 for short runs, suggesting one restrictive mistake and one parsimony mistake on average. These improve to WFM scores of 73.9-83.0 in long runs, suggesting higher accuracy with longer time-series if the reporting patterns are consistent over time. 

In settings with identifiable measurement error (DS0-DS3B), we would prefer to model the measurement error approach rather than smooth over it. This preference holds even in settings where there is no measurement error (DS0-DS1B), since $\boldsymbol{\theta}=\mathbf{1}$ is within the solution set of our model. In settings when measurement error in case-reporting cannot be separated from transmission signals (DS3C), we would prefer to reduce measurement error with the case-smoothing approach. Our decision tool works well in our simulations and rarely selected the case-smoothing approach in DS0-DS3B (0.0-3.3\%) but always did in DS3C.

Finally, estimates of $R_t$ when reporting periods are estimated using Algorithm~\ref{alg:dow_alg} and the measurement error approach is selected using the decision tool demonstrated improved stability across scenarios. Estimation of the reporting periods expectedly reduced $R_t$ estimate efficiency compared to when reporting periods were given a priori, particularly in simple scenarios. However, use of the decision tool greatly improved estimate accuracy in  DS3C.   

\section{Analysis of COVID-19 outbreak}
\label{sec:application}

\subsection{Study Design}

We demonstrate the proposed method in practice using data from the COVID-19 pandemic. We collected reported cases and measures of social distancing and temperature from March 25, 2020 to November 10, 2020 (230 days) for 10 United States (US) counties that contain large metropolitan cities – Austin, Texas; Chicago, Illinois; Dallas, Texas; Houston, Texas; Los Angeles, California; Miami, Florida; New York, NY; San Diego, California; and San Francisco, California. Although data is collected at the county level, we refer to the major city of the county throughout this section. Our interests are twofold, as we wish to both estimate transmission and explore transmission dynamics, such as the role of social distancing and temperature on transmission.

We used \textit{daily visitation difference}, which measures the relative change in visits to non-essential businesses from pre-pandemic times, and  \textit{wet bulb temperature}, which measures the complex thermodynamic relationship between temperature and humidity, as predictors of transmission. These variables, provided by Unacast and the National Ocean and Atmospheric Administration Local Climatological Data respectively, were previously found to be associated with transmission (\cite{Rubin2020}). We used the average, standardized metrics from days $[t-15, t-2]$ as $\mathbf{X}_t$ and included 4 lag terms ($m=4$) in our autoregressive transmission model. 

In addition to county-specific data on reported cases and transmission factors, we selected serial interval distribution estimates from China (\cite{He2020}), Ireland (\cite{McAloonIrelandSI2021}), and a systematic review on the world population (\cite{AleneSerialInterval2021}) as components of our serial interval mixture. The study by \cite{He2020} was used for a study on California counties (\cite{WordenCalifornia2020}), whereas the latter two provide comprehensive studies from different parts of Europe and Asia. The lack of available serial interval studies on a diverse US population highlights the practical challenges that practitioners often face when searching for a representative serial interval study. While some of these studies account for the possibility that a generated case is identified prior to that of the generator, we did not and instead adjusted their estimated serial interval distributions to maintain the relative densities over the positive support of the distribution. For the baseline method implementation, we used the average of these three serial interval estimates.

\subsection{Transmission Estimation}

\begin{figure}
\centering
    \includegraphics[width=5in,keepaspectratio]{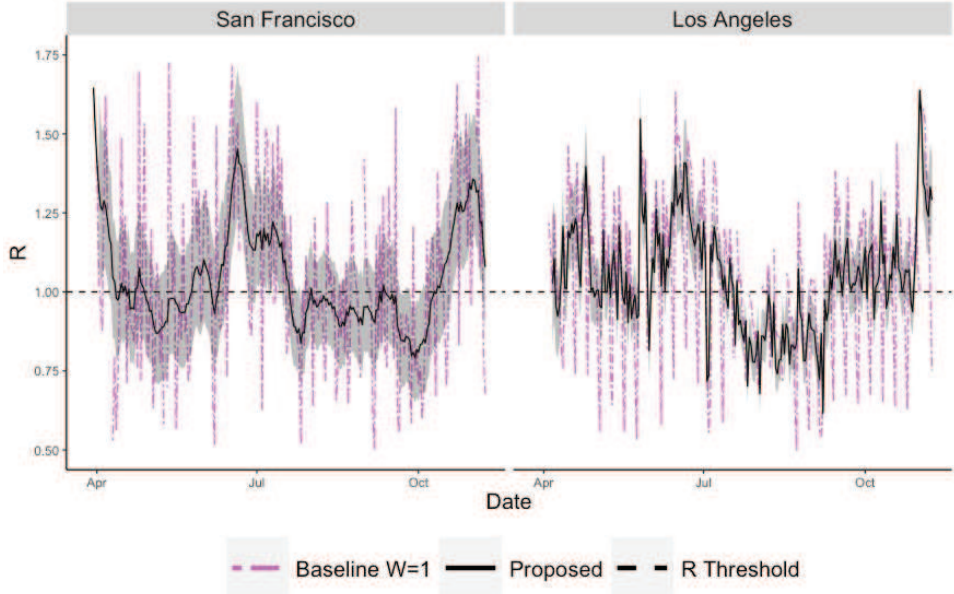}
	\caption{Comparisons of $\hat{R}$ for San Francisco (left) and Los Angeles (right) from March 25, 2020 to November 11, 2020 using the baseline and proposed methods.}
	\label{fig:application_fit_comp}
\end{figure}

We implemented methods as in Section~\ref{sec:simulations}, except we used 12K MCMC samples (6K burn-in) per chain to account for the larger parameter space of a 230-day study. We estimated reporting periods for each county using Algorithm~\ref{alg:dow_alg} and selected an appropriate measurement error approach using the case-smoothing decision tool. Our decision tool selected pre-analysis case-smoothing for two of the counties (Houston and Austin) and the case reporting measurement error model for the other eight. 

Figure \ref{fig:application_fit_comp} visualizes $R_t$ estimates from the proposed and baseline $W=1$ methods for both San Francisco and Los Angeles. Qualitatively, the proposed method produced smoother and more usable estimates than the baseline method. Importantly, this smoothing is conducted in a more principled fashion than window-based smoothing techniques since it directly targets nuisance reporting patterns. 

\begin{figure}
	\centering
	\includegraphics[width=5.5in]{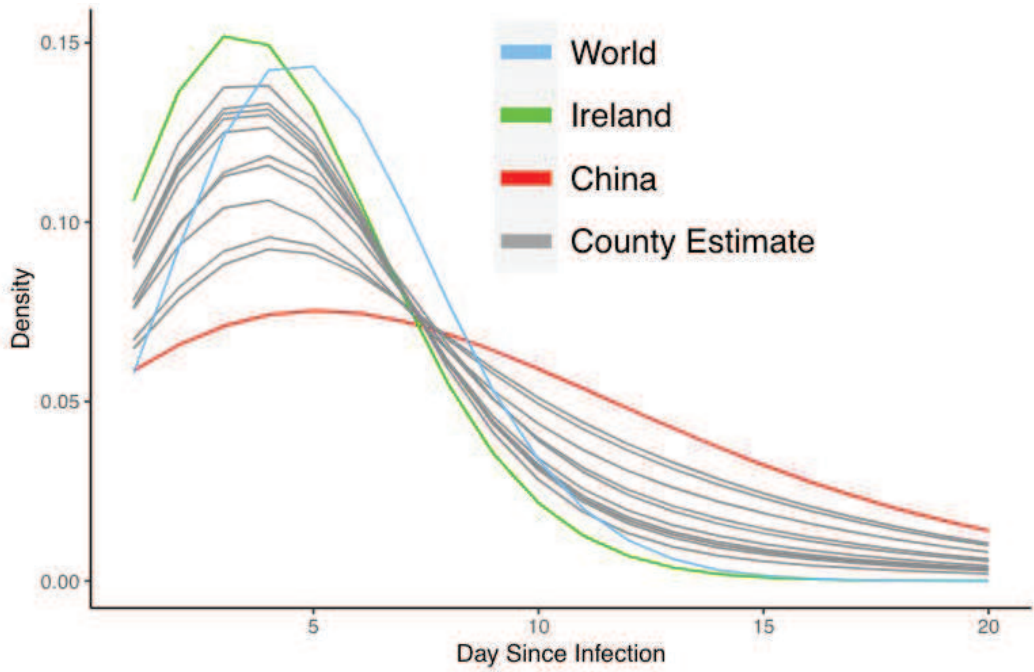}
	\caption{A comparison of the external study serial intervals used and the fitted serial intervals for the counties in the study.}
	\label{fig:county_serial_intervals}
\end{figure}

While we suspect that much of the improved estimate smoothness results from the identification of case-reporting patterns, it may also result from increases in estimate stability brought on by our transmission predictor model and/or flexibility brought on by allowing deviation from a specified serial interval distribution. To the latter point, Figure~\ref{fig:county_serial_intervals} shows heterogeneity in serial interval distribution estimates for each county, suggesting that such variations enabled a better overall model fit. Interestingly, Texas and California had the most and least volatile $R_t$ estimates, respectively, which may suggest more identifiable reporting patterns in the latter.

\subsection{Exploring Factors of Transmission}

\begin{figure}
\centering
	\includegraphics[width=5.5in]{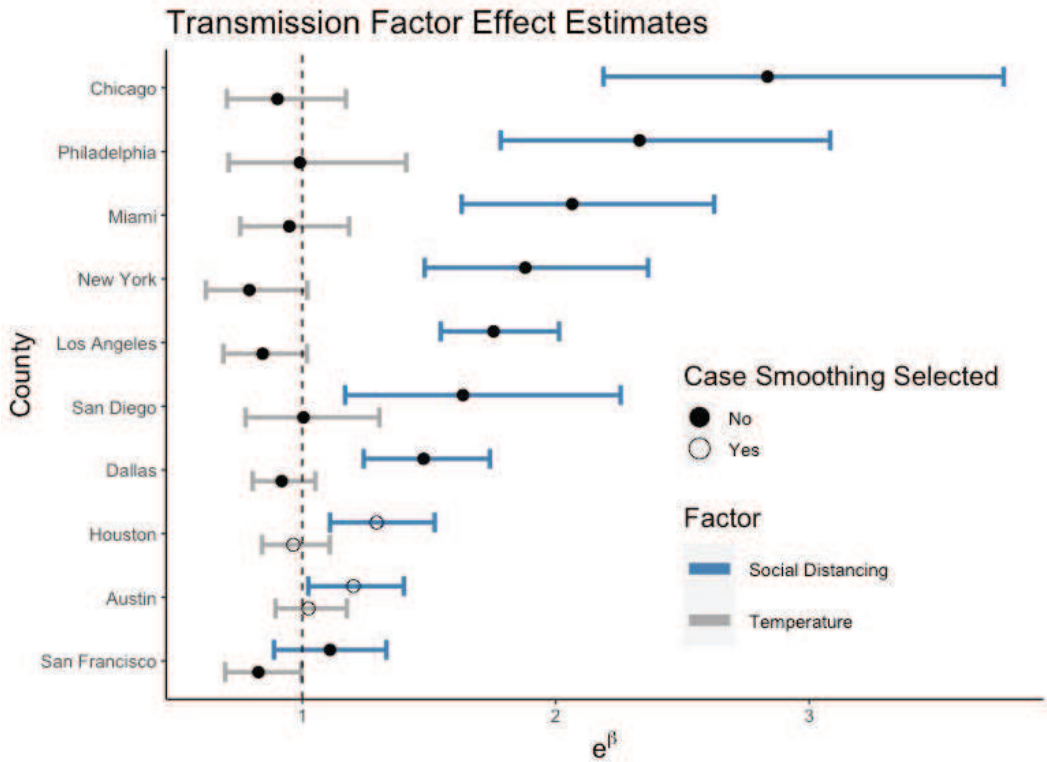}
	\caption{$\beta$ estimates from the proposed model for social distancing (blue) and wet bulb temperature (gray). }
	\label{fig:ottoman_tables}
\end{figure}

\begin{figure}
	\centering
	\includegraphics[width=5.5in]{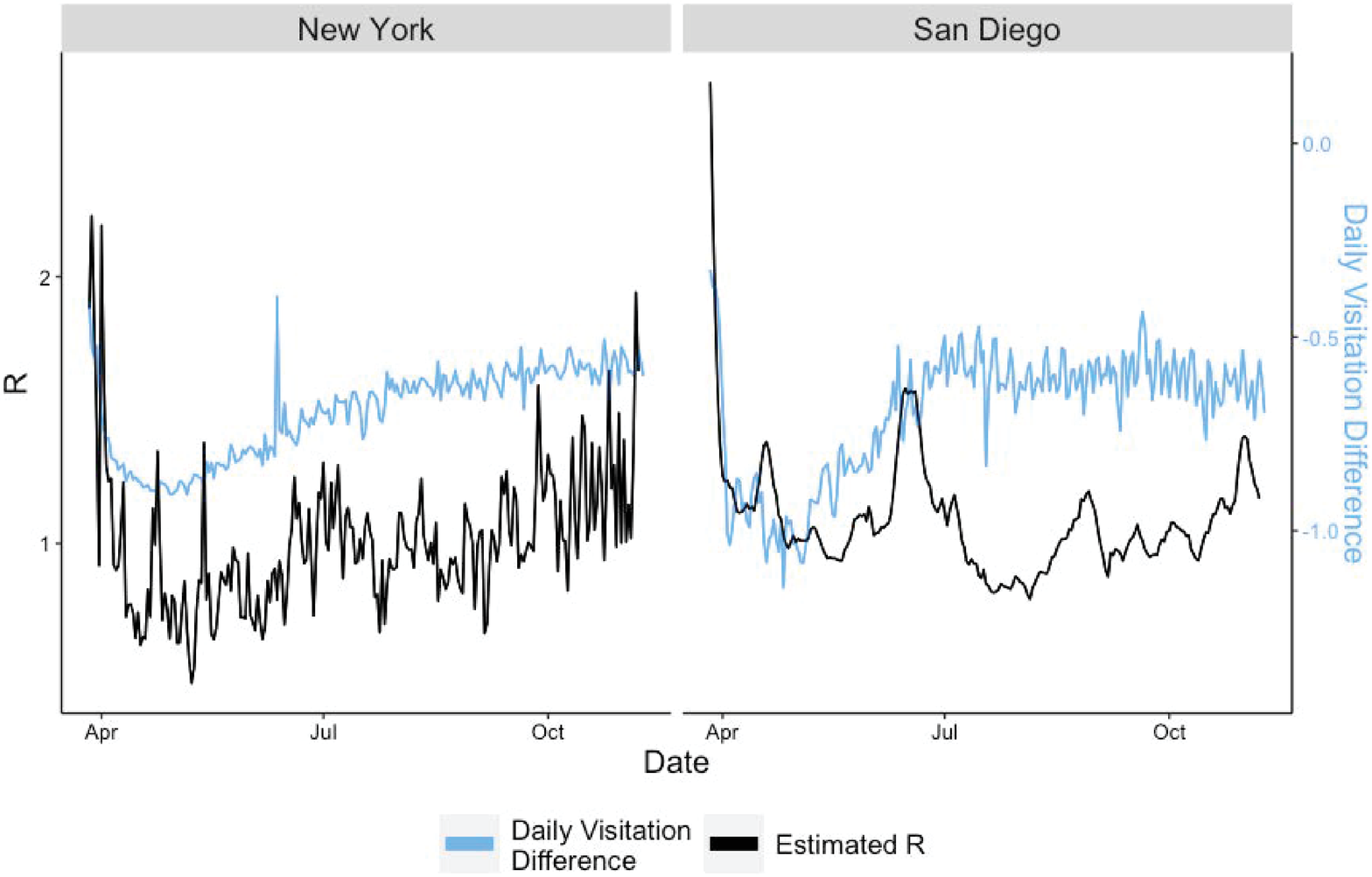}
	\caption{A comparison of the estimated reproduction number and rolling averages of daily visitation difference for New York (left) and San Diego (right).}
	\label{fig:full_transmission_soc}
\end{figure}

We explored the association of specified covariates and transmission by examining the $\boldsymbol{\beta}$ coefficient estimates from our method, which are shown in Figure~\ref{fig:ottoman_tables}. The association between social distancing and transmission for San Diego and New York is demonstrated qualitatively in Figure~\ref{fig:full_transmission_soc}. Each additional unit (county-specific standard deviation) of daily visitation difference is associated with an $100(e^\beta-1)\%$ increase in expected $R_t$. Thus, our estimates associate an increase of $9-244\%$ in expected $R_t$ per standard deviation increase in social distancing relative to pre-pandemic baselines. The confidence intervals for these associations do not include the null effect ($e^\beta=1$) except for San Francisco, which has highly volatile and often even positive daily visitation differences in the data. 

Our estimates for the effect of wet bulb temperature showed a negative but inconclusive association with the transmission, which differs from previous studies that found significant negative associations (\cite{Rubin2020}). The high correlation of temperature with the presumably stronger and offsetting social distancing measure ($r=0.53 - 0.85$) and the suspected non-linear relationship between temperature and transmission may partially explain the weakened association. 

\section{Discussion}
\label{sec:discussion}

When estimating $R_t$, practitioners must use noisy case-notification data subject to reporting patterns and non-representative external serial interval estimates. In this work, we demonstrated the importance of modeling case reporting measurement error and accounting for serial interval uncertainty when estimating $R_t$ with imperfect data and proposed a method to do so. Our Bayesian framework shares information and uncertainty between intuitive modeling layers, allowing for informed but flexible parameter exploration. In our simulation studies, these benefits enabled more robust parameter and uncertainty estimates than methods that do not target these practical considerations. Further, our objective methods to determine the case reporting measurement error approach augment our proposed model and equip practitioners with tools to avoid ad-hoc smoothing selection techniques. We applied our framework to a study on the 2020 US COVID-19 outbreak to derive potentially more usable $R_t$ estimates. In addition to benefiting monitoring efforts, these estimates may better parameterize models that rely on $R_t$ as we avoid smoothing over relevant signals in the data. Finally, our framework allowed us to explore the association of transmission factors in explaining transmission, which suggested a large influence of social distancing on transmission.

Our framework targets measurement error due to case reporting and can be viewed as a new approach to smoothing, where our smoothing technique incorporates our knowledge of differential reporting behavior across days of the week. When day-of-week reporting variation does not sufficiently capture overall reporting variation, coarser methods like rolling averages or spline-smoothing may be beneficial. Our approach does not address unidentified cases, which result from factors like asymptomatic transmission and testing guidelines. When the rate of case-identification is constant over time, day-of-week reporting weights ($\boldsymbol{\theta}$) are identifiable up to a constant and our estimates for $R_t$ are unbiased as this constant rate is canceled out in the renewal equation. However, $R_t$ estimates are biased if the identification rate changes over time (\cite{WhiteReporting2010}) and while methods have been developed to use additional data sources to account for these concerns (\cite{WhiteInfluenza2009, QuickLin2021}), they come with their own assumptions about additional data and modeling layers (\cite{DeanCommentQuick2021}). Changes in reporting patterns may also trigger changes in the serial interval over time (\cite{DattaCommentQuick2021, QuickRejoinder2021}). In addition to different sources of data, methods that integrate case data from multiple regions into the estimation process may provide benefits, but also come with substantial challenges.

While our approach provides insights relevant to future transmission and the effects of transmission factors, it is designed to estimate $R_t$ rather than to forecast future transmission or robustly analyze effects. Nevertheless, recent works have demonstrated the predictive power of autoregressive transmission models in practice suggesting similar methodology may perform well for forecasting tasks (\cite{DouwesSchultz2022ExtendedForecasting, Shi2022RobustSARS-CoV-2}). When analyzing the effects of policies or transmission factors, \cite{LipsitchPanComment2020} advocate for using the case reproduction number, $R^c_t$, as this measure is estimated using additional data after day $t$. We further caution against strong interpretation of the associations between predictors and outcomes, as the predictors are used in the estimation of $R_t$, which may induce bias to larger effect sizes. Still, the relative strength of associations, particularly for standardized covariates as demonstrated in our study, may provide insights into the relative importance of covariates. 

Finally, our current implementation relies on several parametric assumptions. While it is straightforward to implement our model using distributions with additional parameters to allow for features like over-dispersion, semi- or non-parametric approaches would allow further flexibility. \cite{Shi2022RobustSARS-CoV-2} use a novel quasi-score approach to reduce parametric reliance and it would be valuable to understand if extensions to their model would be as seamless as in a Bayesian hierarchical framework.

\section*{Available Code}

Code and an example simulated dataset are provided on GitHub at https://github.com/garyhettinger/TransmissionEstimator.
\vspace*{-8pt}

\bibliographystyle{unsrtnat}
\bibliography{references}

\appendix

\setcounter{equation}{0}
\setcounter{figure}{0}
\setcounter{table}{0}
\setcounter{section}{0}
\renewcommand{\theequation}{A\arabic{equation}}
\renewcommand{\thefigure}{A\arabic{figure}}
\renewcommand{\thesection}{A\arabic{section}}
\renewcommand{\thetable}{A\arabic{table}}
\renewcommand{\thefigure}{A\arabic{figure}}

\label{sec:appendix}

\section{Constraints on and Identifiability of Reporting Weights and Reported Cases}

When addressing measurement error in case reporting caused by patterns like batched reporting rather than unidentified cases as we do in our method, it may seem reasonable to impose some constraint on the relationship between observed cases, $I_t$, and the measurement error-free cases, $I^*_t$. For example, imposing $\sum\limits_t I_t = \sum\limits_t I^*_t$ would ensure that our method only estimates a different time for the reporting of observed cases rather than adding or removing any reported cases in aggregate. However, such constraints can be difficult to build into a model. 

Under our proposed case reporting measurement error model where $\{\tau^* : \tau(t) = \tau^*\}$, we have
\begin{equation*}
    E[I_t] = \theta_{\tau(t)} I^*_t \implies \sum\limits_t E[I_t] = \sum\limits_{\tau^*} \theta_{\tau^*} \sum\limits_{t^*: \tau(t^*) = \tau^*} I^*_{t^*}
\end{equation*}
Under our Dirichlet prior, we then have
\begin{equation*}
    \sum\limits_{t^*_1: \tau(t^*_1) = \tau^*_1} I^*_{t^*_1} = \sum\limits_{t^*_2: \tau(t^*_2) = \tau^*_2} I^*_{t^*_2} \hspace{3mm} \forall \tau^*_1 \neq \tau^*_2 \implies \sum\limits_t E[I_t] = \sum\limits_t I^*_t
\end{equation*}

Having similar total true case counts across reporting periods may be an unreasonable expectation in practice, particularly with unbalanced periods (e.g., 3 days-of-week in one group and 4 in the other). Still, this constraint further helps with $\boldsymbol{\theta}$, and thus $R_t$, identifiability, even in cases when assumptions of the Dirichlet prior are violated. This observation results from the partial identifiability of $\boldsymbol{\theta}$ up to a constant, $c$:
\begin{equation*}
    R_t = \frac{E[I^*_t]}{\sum\limits_s w^*_s I^*_{t-s}} = \frac{E[I_t] / \theta_{\tau(t)}}{\sum\limits_s w^*_s E[I_{t-s}]/\theta_{\tau(t-s)}} = \frac{E[I_t] / (c\theta^*_{\tau(t)})}{\sum\limits_s w^*_s E[I_{t-s}]/(c\theta^*_{\tau(t-s)})} = \frac{E[I_t] / \theta^*_{\tau(t)}}{\sum\limits_s w^*_s E[I_{t-s}]/\theta^*_{\tau(t-s)}}
\end{equation*}
where we use $E[I_t] = \theta_{\tau(t)} I^*_t$ and $E[I_t] = R_t \sum\limits_s w^*_s I^*_{t-s}$ per our model assumptions. Thus, the Dirichlet prior may actually help identifiability by bounding the mean of $\boldsymbol{\theta}$. This observation was noted in our strong identification of $R_t$ even when the Dirichlet prior was violated in DS3B.

\section{Simulation Parameter Values}

In DS1B, DS2A-DS3C, the GARMA model for $R_t$ is specified with AR-order of 2 and MA-order of 0, along with auto-regression coefficients $\boldsymbol{\phi}=(0.4, -0.167)$. In DS0, DS1A, and DS1C, AR-order is 0. For all scenarios, we specified covariate coefficients as $\boldsymbol{\beta} = \begin{cases} (1.21, 2.24) & TP \neq 3 \\ (1.10,1.82) & TP=3 \end{cases}$

In DS0-DS3B, serial interval estimates are specified as $\mathbf{w^{(1)}} = (0.8, 0.1, 0.075, 0.025)$, $\mathbf{w^{(2)}} = (0.1, 0.4, 0.3, 0.2)$, and $\mathbf{w^{(3)}} = (0.05, 0.15, 0.15, 0.65)$, with $\boldsymbol{\lambda} = (0.1, 0.7, 0.2)$. In scenarios where the serial interval is given (DS0, DS1B, DS1C, DS3C), both baseline and proposed models are given $\mathbf{w^*} = \lambda_1 \mathbf{w^{(1)}} + \lambda_2 \mathbf{w^{(2)}} + \lambda_3 \mathbf{w^{(3)}}$. Otherwise, the proposed model is given $\mathbf{w^{(1)}}$, $\mathbf{w^{(2)}}$, and $\mathbf{w^{(3)}}$ whereas the baseline model is passed $\mathbf{\bar{w}} = (\mathbf{w^{(1)}} + \mathbf{w^{(2)}} + \mathbf{w^{(3)}})/3$. In DS3C, $\mathbf{w^*} = (1,0,0,0)$, enforcing that all infections are generated by the previous day. 

In DS0-DS1B, reporting weights are specified as $\boldsymbol{\theta} = \mathbf{1}$. In DS1C-DS2B, $\boldsymbol{\theta} = (0.5, 1.5, 1.0)$ and $\tau(t) = \{(M, Tu), (W, Th, F), (Sa, Su)\}$ In DS3A, $\boldsymbol{\theta} = (0.5, 1.5, 1.0)$ in the first half of the outbreak and then $\boldsymbol{\theta} = \mathbf{1}$ in the second half of the outbreak. In DS3B, $\boldsymbol{\theta} = (0.5, 1.25, 0.8)$. Finally, in DS3C, $\boldsymbol{\theta} = \begin{cases} (0.5, 1.23) & \text{unrestricted phase} \\ (0.5, 2) & \text{lockdown phase}\\ (0.5, 1.5) & \text{equilibrium phase} \end{cases}$

\section{Additional Simulation Metric Summaries}

In Table A3, we break out the performance of our proposed method summarized in Table 1 by by simulation trend pattern. We generally see the best performance for TP3 simulations, where mean transmission decays smoothly, and worst performance for long simulations. We suspect these observations are largely influenced by less (and more) volatile case peaks introduced in the two simulation settings. 

Performance metrics of additional key parameter estimates for DS2A are provided in Table A4. These metrics suggest reasonable identification when models are well-specified. The relatively high MSE and bias of $\boldsymbol{\lambda}$ estimates does not seem to negatively affect estimation of $\mathbf{w^*}$. This may suggest that distinct $\boldsymbol{\lambda}$ parameters create similar serial interval distributions in our relatively simple simulation scenarios with a maximum infection time of 4 days.

\vspace{5in}

\section{Supplemental Figures}

\begin{figure}[H]
     \centering
     \begin{subfigure}[b]{0.48\textwidth}
         \centering
         \includegraphics[width=\textwidth]{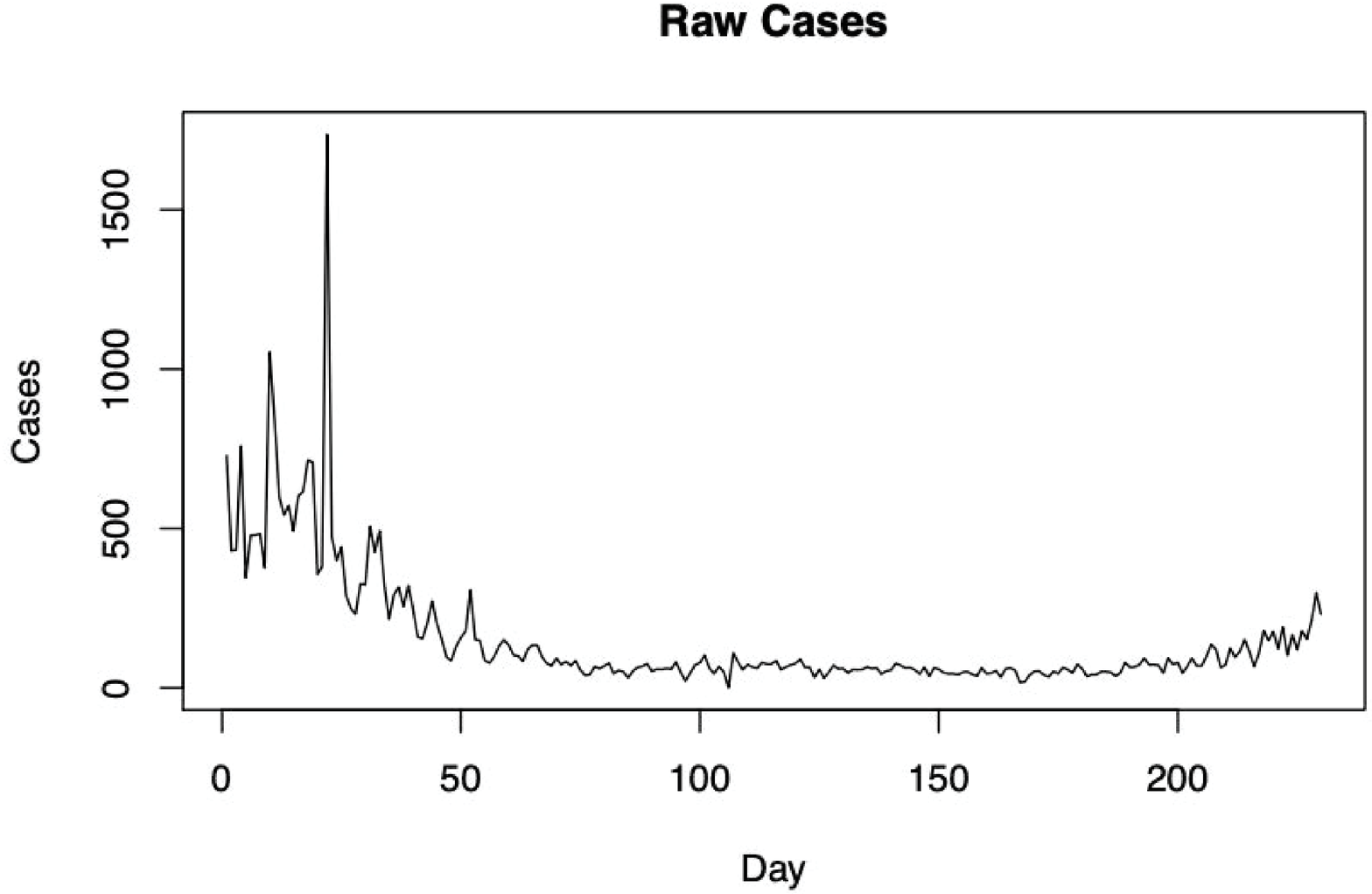}
         \caption{Begin with the reported cases, $I_t$.}
         \label{fig:me_model_process:0}
     \end{subfigure}
     \hfill
     \begin{subfigure}[b]{0.48\textwidth}
         \centering
         \includegraphics[width=\textwidth]{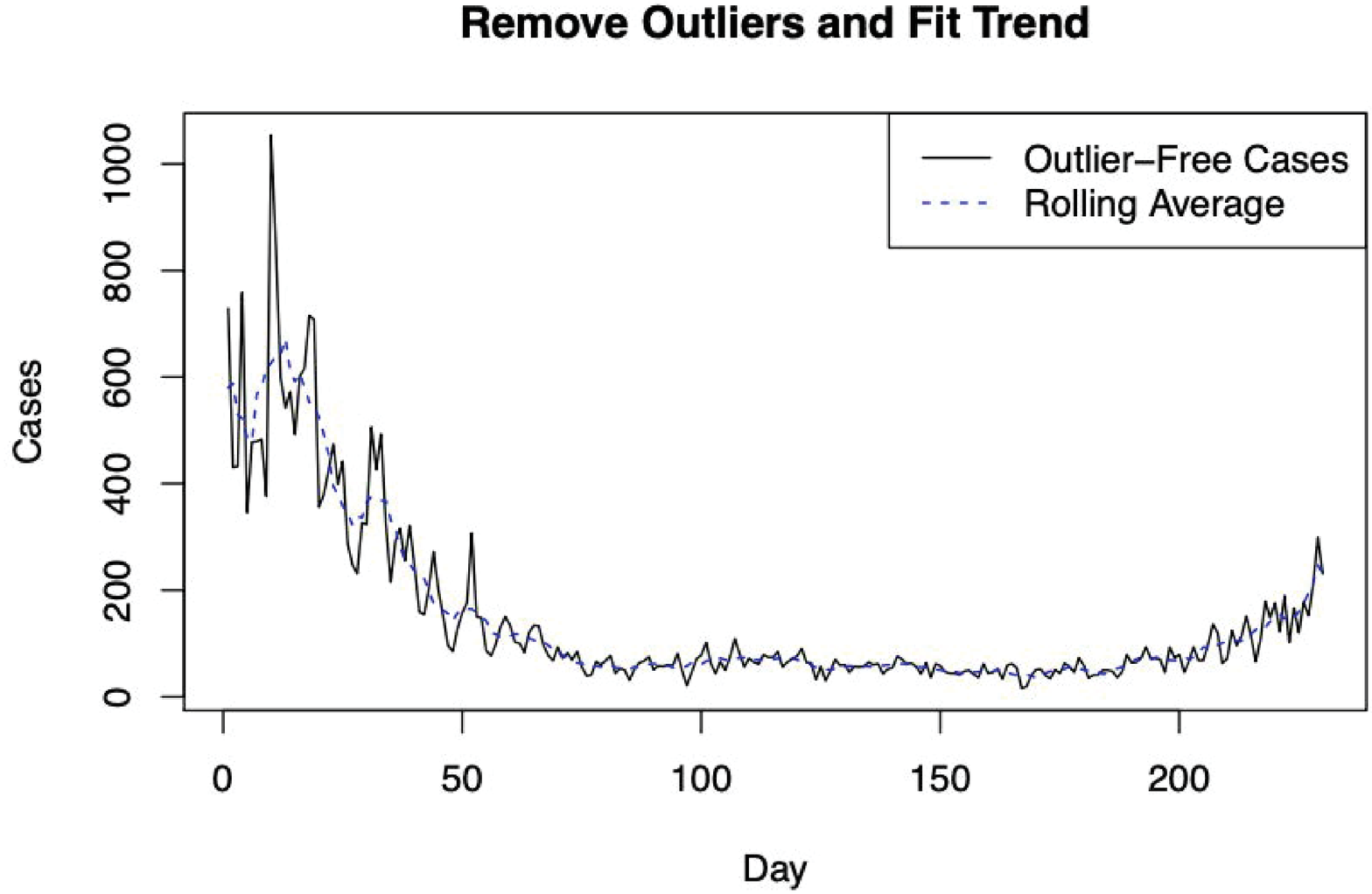}
         \caption{Compute rolling average as a proxy for $I^*_t$.}
         \label{fig:me_model_process:1}
     \end{subfigure}
     \begin{subfigure}[b]{0.48\textwidth}
         \centering
         \includegraphics[width=\textwidth]{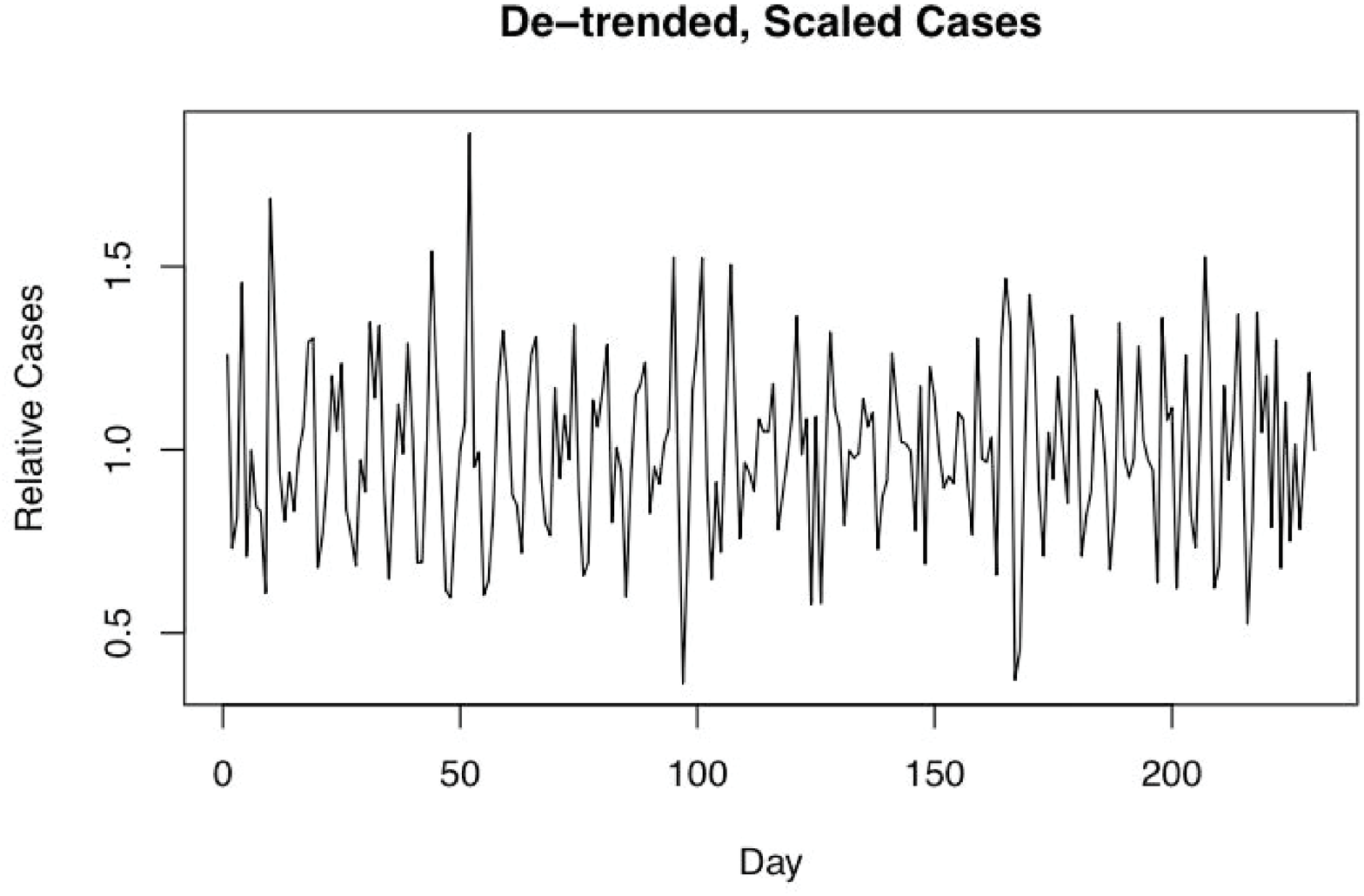}
         \caption{Then, detrend and scale cases as a proxy for relative reporting variation, $V_t$.}
         \label{fig:me_model_process:2}
     \end{subfigure}
     \hfill
     \begin{subfigure}[b]{0.48\textwidth}
         \centering
         \includegraphics[width=\textwidth]{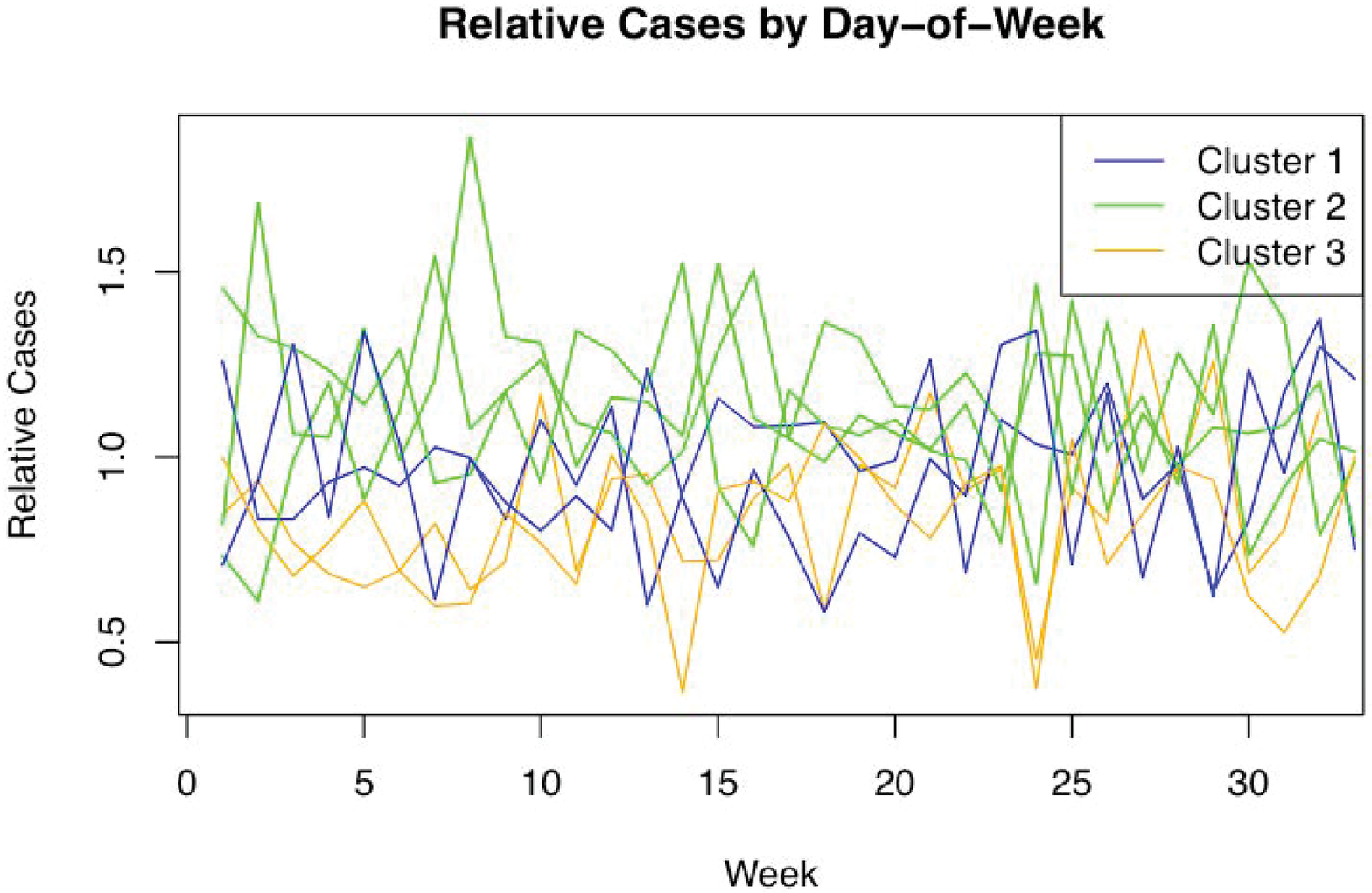}
         \caption{Separate reporting variation by day-of-week,  $v^{(l)}$, which shows some heterogeneity.}
         \label{fig:me_model_process:3}
     \end{subfigure}
     \begin{subfigure}[b]{0.6\textwidth}
         \centering
         \includegraphics[width=\textwidth]{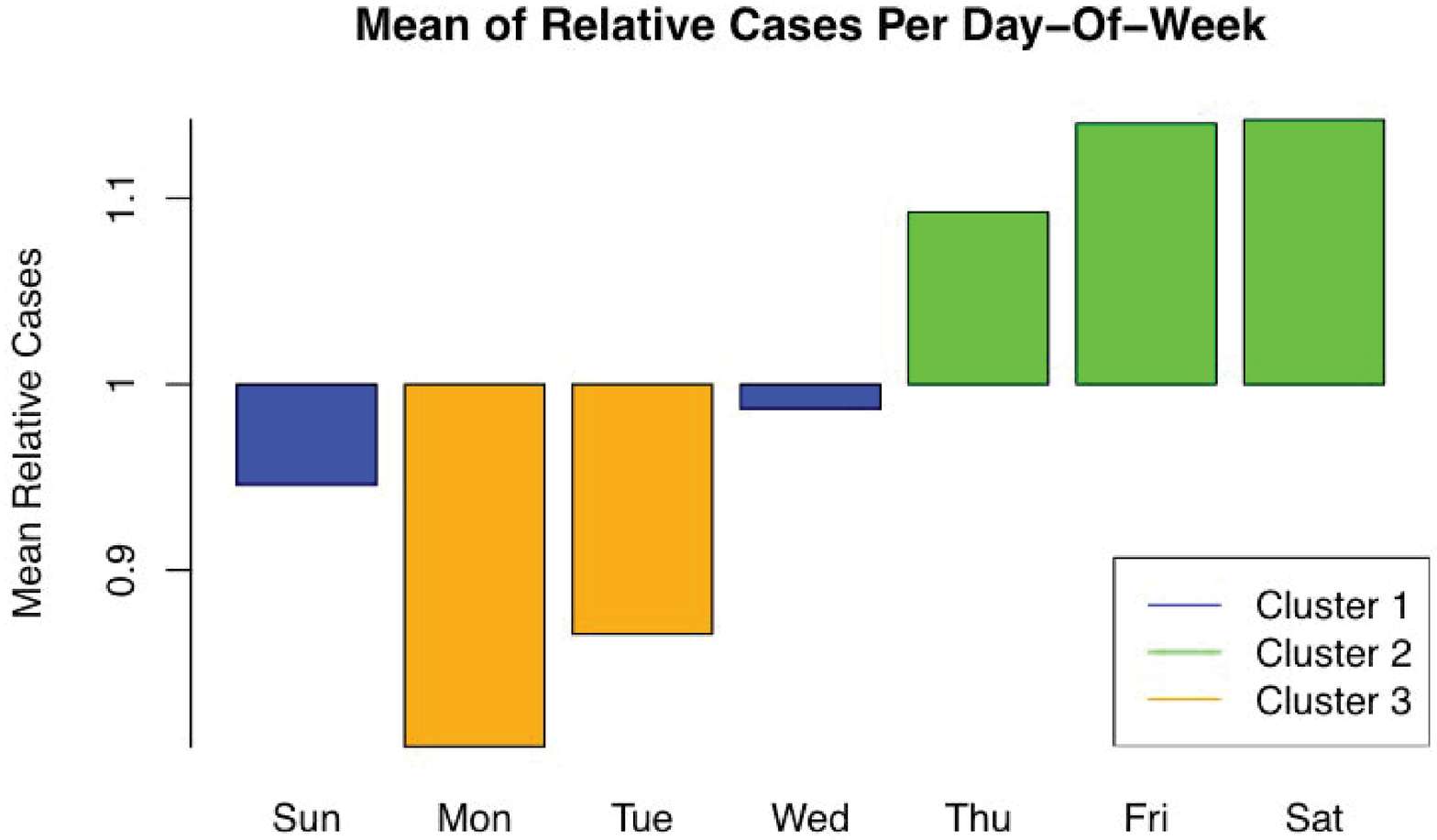}
         \caption{Select clustering that minimizes AIC.}
         \label{fig:me_model_process:4}
     \end{subfigure}
        \caption{Visualization of Algorithm 1 using COVID-19 case data from New York.}
        \label{fig:me_model_process}
\end{figure}

\begin{figure}[H]
	\centering
	\includegraphics[width=6cm,height=6cm,keepaspectratio]{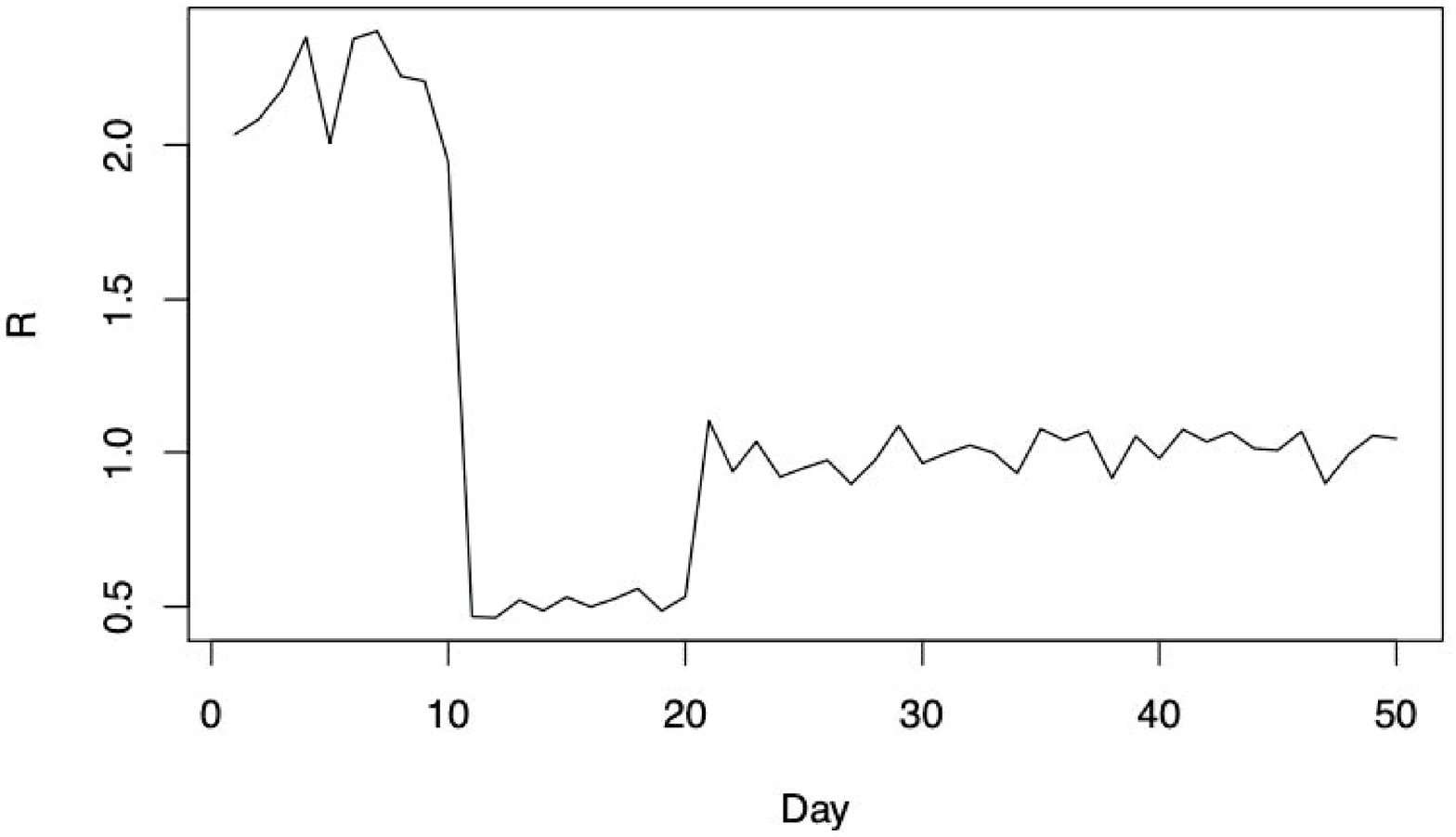}
	\includegraphics[width=6cm,height=6cm,keepaspectratio]{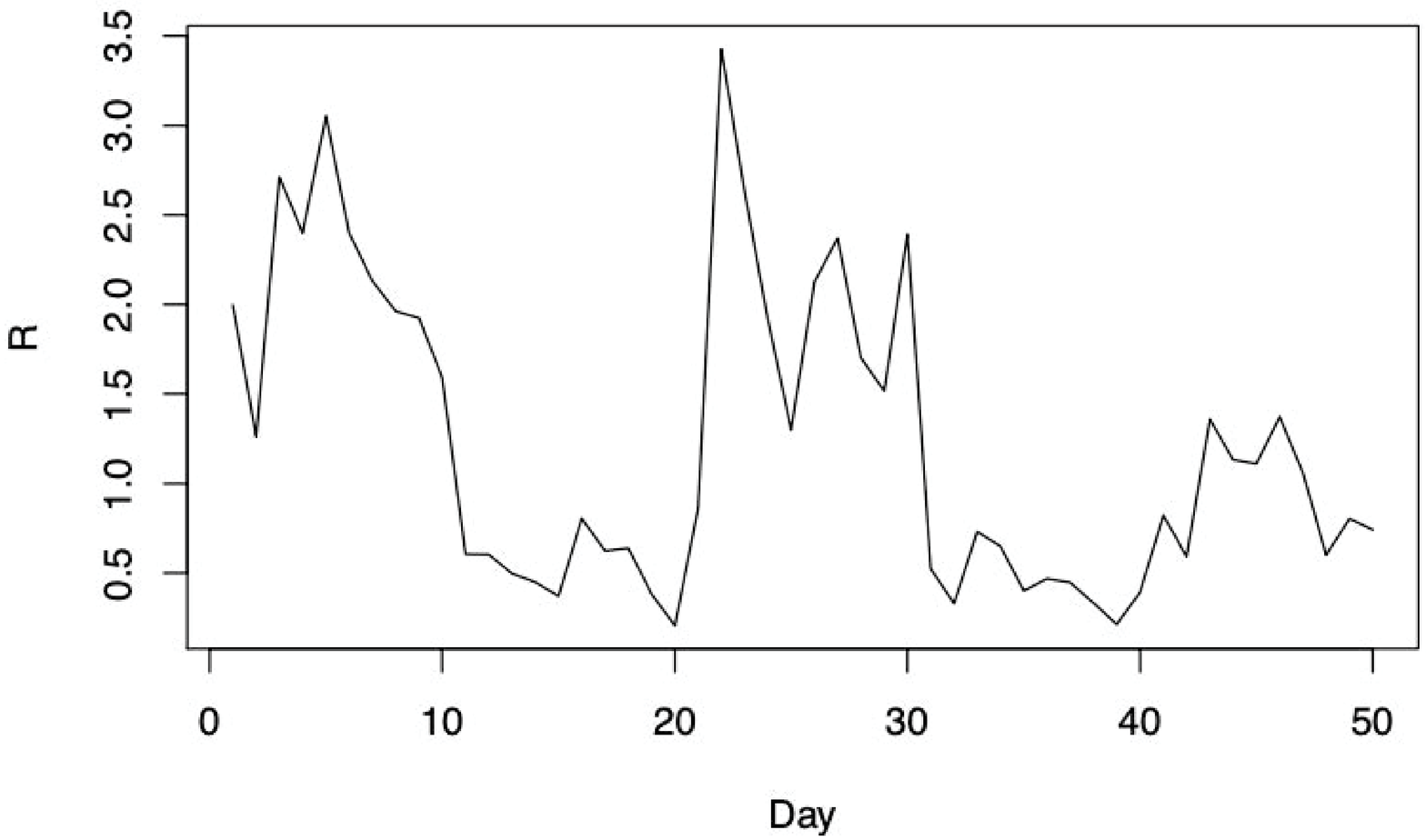}
	\includegraphics[width=6cm,height=6cm,keepaspectratio]{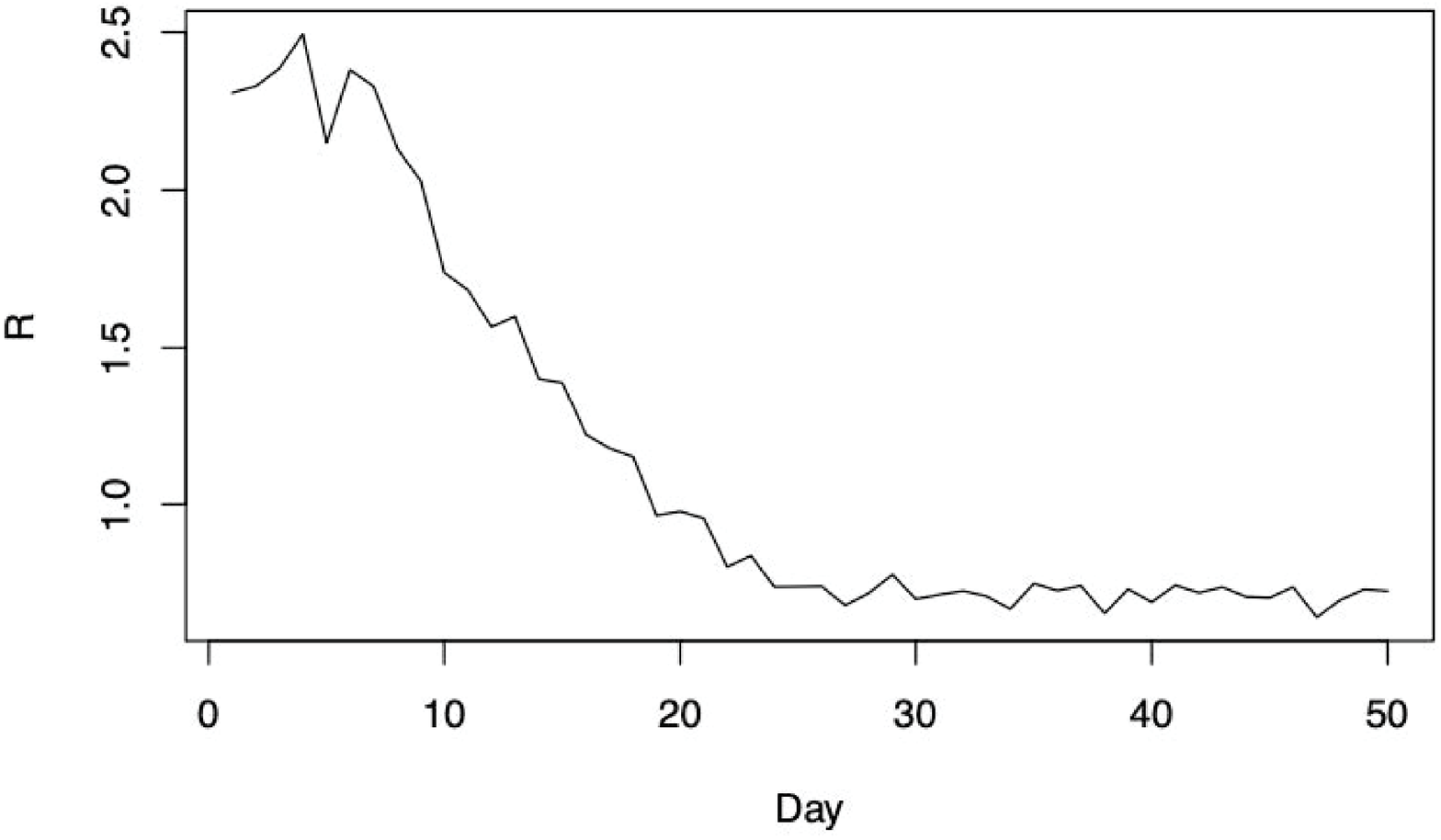}
	\caption{Simulated $R_t$ from trend patterns 1 (Top Left), 2 (Top Right), and 3 (Bottom)}
	\label{fig:contexts}
    \vspace{-6.5mm}
\end{figure}

\begin{figure}[H]
     \centering
     \begin{subfigure}[b]{0.48\textwidth}
         \centering
         \includegraphics[width=\textwidth]{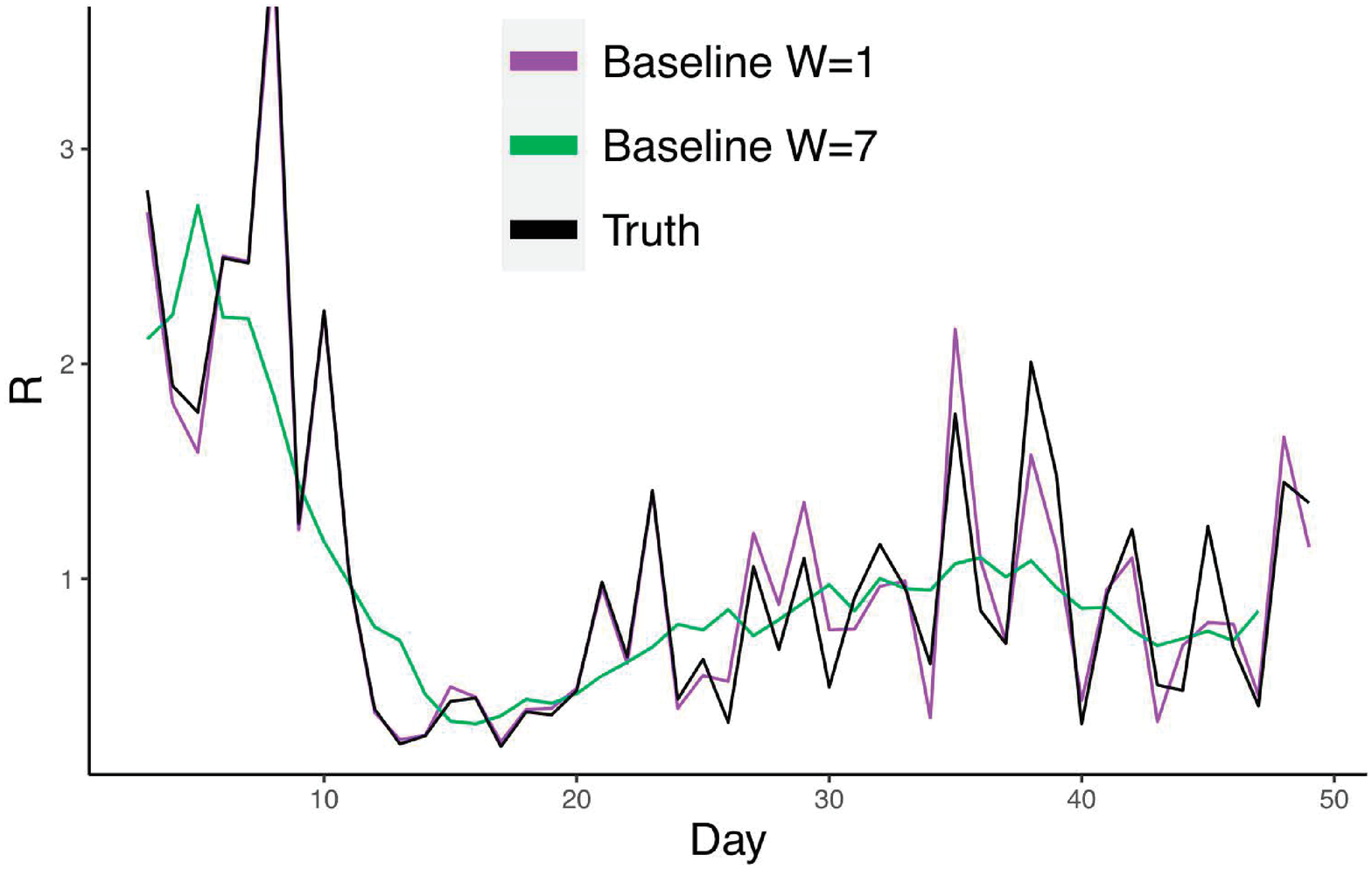}
         \caption{DS0}
         \label{fig:fits:scen0}
     \end{subfigure}
     \hfill
     \begin{subfigure}[b]{0.48\textwidth}
         \centering
         \includegraphics[width=\textwidth]{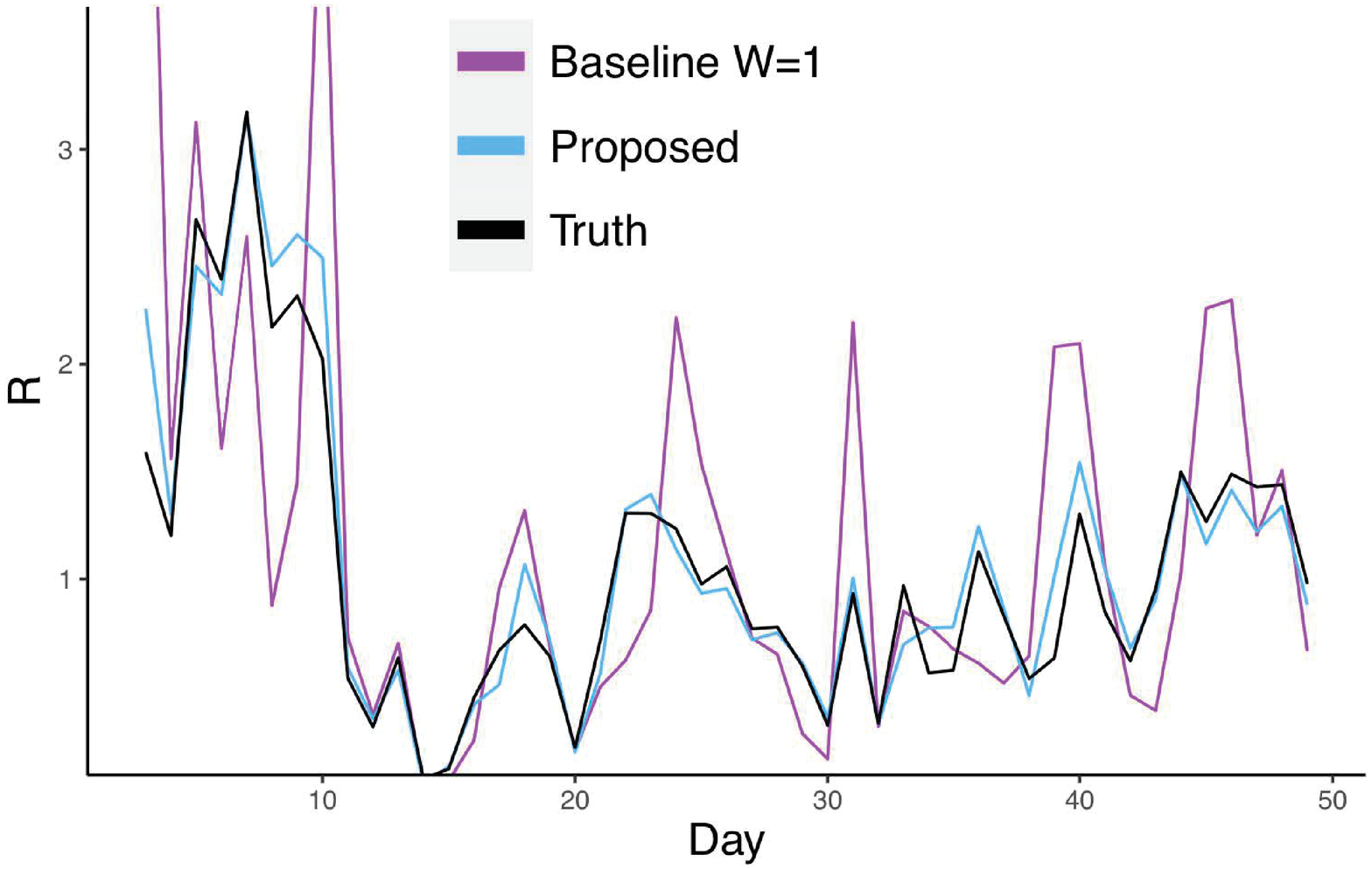}
         \caption{DS2A}
         \label{fig:fits:scen2a}
     \end{subfigure}
     \hfill
     \begin{subfigure}[b]{0.6\textwidth}
         \centering
         \includegraphics[width=\textwidth]{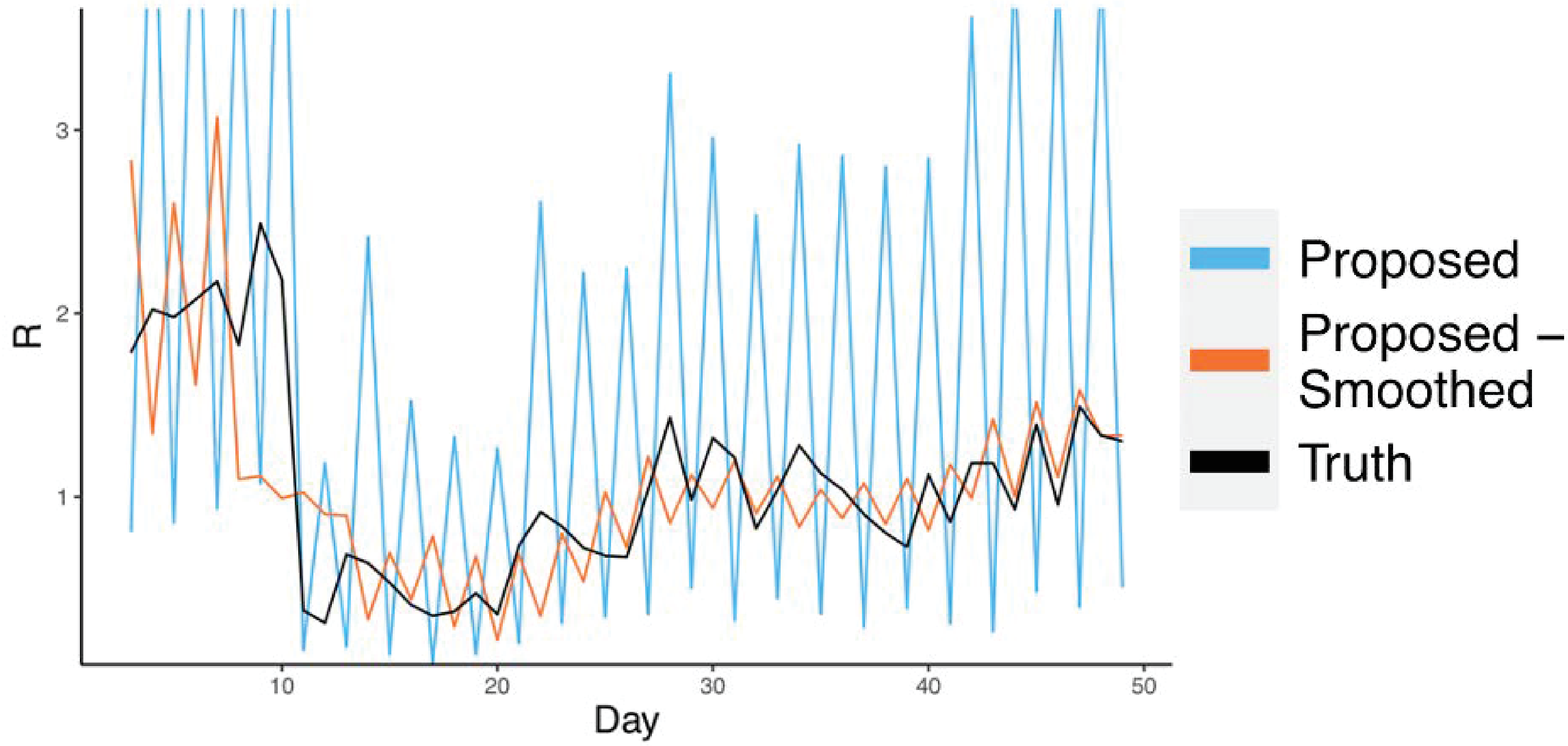}
         \caption{DS3C}
         \label{fig:fits:scen3c}
     \end{subfigure}
        \caption{Example simulations with model fits. (a) Example from an outbreak with accurate case reporting and a correctly-specified serial interval distribution (DS0) comparing the effect of window size (W) on baseline model estimates. (b) Example from an outbreak with an identifiable reporting pattern, mixture serial interval distribution, and informative covariates (DS2A) comparing proposed model fit to baseline model fit. (c) Example from an outbreak with an unidentifiable reporting pattern (DS3C) comparing the proposed method with a case reporting measurement error model and with pre-analysis case-smoothing.}
        \label{fig:fits}
\end{figure}

\section{Supplemental Tables}

\begin{table}[H]
	\caption{Performance of $R_t$ estimates for baseline method by window length (W) and serial interval uncertainty (U).}
	\centering
	\begin{tabular}{l|lll|lll|lll}
		\hline
		Data & \multicolumn{3}{c}{MSE (x$10^{-2}$)} & \multicolumn{3}{c}{Bias ($\%$)} & \multicolumn{3}{c}{Coverage Probability ($\%$)} \\
		\cline{2-4} \cline{5-7} \cline{8-10}
		Scenario & W=1 (U) & W=1 & W=7 & W=1 (U) & W=1 & W=7 & W=1 (U) & W=1 & W=7 \\
		\hline
		0 & -$^*$ & \textbf{1.44} & 35.82 & - & \textbf{2.01} & 36.13 & - & \textbf{95.13} & 10.29\\
        1a & 6.25 & \textbf{2.62} & 36.23 & 8.19 & \textbf{1.93} & 36.05 & \textbf{86.66} & 70.65 & 10.16\\
        1b & - & \textbf{2.06} & 41.61 & - & \textbf{3.20} & 45.07 & - & \textbf{95.07} & 10.44\\
        1c & - & 41.94 & \textbf{36.80} & - & \textbf{10.34} & 36.22 & - & \textbf{16.86} & 7.57\\
        2a & 82.45 & \textbf{39.23} & 44.61 & 27.96 & \textbf{10.15} & 48.97 & \textbf{33.91} & 19.80 & 8.14\\
        2b & 92.45 & 52.82 & \textbf{14.29} & 22.98 & 8.31 & \textbf{6.40} & \textbf{30.31} & 16.25 & 12.51\\
        3a & 46.61 & \textbf{24.40} & 43.56 & 17.91 & \textbf{5.69} & 49.12 & \textbf{54.82} & 37.85 & 8.31\\
        3b & 60.44 & \textbf{31.01} & 43.61 & 22.97 & \textbf{8.62} & 47.28 & \textbf{37.39} & 22.61 & 8.58\\
        3c & - & 272.30 & \textbf{18.68} & - & 78.27 & \textbf{8.69} & - & 0.72 & \textbf{10.15}\\
		\hline
	\end{tabular}

        {\footnotesize $^*$Baseline models with serial interval uncertainty are only estimated for scenarios when serial interval is unknown.}
	\label{tab:coricomp}
\end{table}

\begin{table}[H]
	\caption{Examples of Weighted F-Measure (WFM).}
	\centering
	\begin{tabular}{l|l|l|ll}
		\hline
		Data Scenarios & True Clustering & Estimated Clustering & No. Mistakes & WFM ($\%$) \\
		\hline
		0,1a,1b & $(1,1,1,1,1,1,1)$ & $(1,1,1,1,1,1,1)$ & 0 & 100 \\
	    0,1a,1b & $(1,1,1,1,1,1,1)$ & $(1,1,1,1,1,2,2)$ & 2 & 84.6 \\
		0,1a,1b & $(1,1,1,1,1,1,1)$ & $(1,1,1,1,1,2,3)$ & 2 & 82.0 \\
        0,1a,1b & $(1,1,1,1,1,1,1)$ & $(1,1,1,1,1,2,3)$ & 2 & 82.0 \\
		0,1a,1b & $(1,1,1,1,1,1,1)$ & $(1,1,1,1,2,3,4)$ & 3 & 66.7 \\
		0,1a,1b & $(1,1,1,1,1,1,1)$ & $(1,1,2,3,4,5,6)$ & 5 & 20.0 \\
		0,1a,1b & $(1,1,1,1,1,1,1)$ & $(1,2,3,4,5,6,7)$ & 6 & 0 \\
		1c, 2, 3, 4b & $(1,1,2,2,3,3,3)$ & $(1,1,2,2,3,3,3)$ & 0 & 100 \\
		1c, 2, 3, 4b & $(1,1,2,2,3,3,3)$ & $(1,1,2,2,3,3,4)$ & 1 & 88.2 \\
		1c, 2, 3, 4b & $(1,1,2,2,3,3,3)$ & $(1,1,2,2,2,3,3)$ & 1 & 60.0 \\
		1c, 2, 3, 4b & $(1,1,2,2,3,3,3)$ & $(1,1,2,2,3,4,5)$ & 2 & 77.0 \\
		1c, 2, 3, 4b & $(1,1,2,2,3,3,3)$ & $(1,1,2,2,2,3,4)$ & 2 & 47.6 \\
		1c, 2, 3, 4b & $(1,1,2,2,3,3,3)$ & $(1,1,2,3,4,5,6)$ & 3 & 55.6 \\
		1c, 2, 3, 4b & $(1,1,2,2,3,3,3)$ & $(1,2,2,3,3,4,5)$ & 4 & 0 \\
		1c, 2, 3, 4b & $(1,1,2,2,3,3,3)$ & $(1,2,3,4,5,6,7)$ & 4 & 0 \\
		\hline
	\end{tabular}

    {\footnotesize $WFM = ((B^2 + 1)*precision*recall) / (B^2*precision+recall)$, where $B$ controls the relative penalization between incorrectly creating an additional cluster (parsimony error) and incorrectly assigning a day to another existing cluster (restrictive error). We set $B=0.5$ to penalize errors of parsimony less than errors of flexibility.}
	\label{tab:example_wfm}
\end{table}

\begin{table}[H]
	\caption{$R_t$ estimation performance for proposed models for simulations with well-specified reporting periods across different simulation trend patterns.}
	\centering
	\begin{tabular}{lll|lll}
		\hline
          Data & Trend & Study & & & Coverage \\
		Scenario  & Pattern & Length & MSE (x$10^{-2}$) & Bias ($\%$) & Probability ($\%$) \\
		\hline
        0 & 1 & Short & 1.28 & 1.63 & 95.03 \\
        0 & 2 & Short & 1.04 & 1.60 & 95.09 \\
        0 & 2 & Long & 2.24 & 3.34 & 94.75 \\
        0 & 3 & Short & 0.60 & 0.96 & 95.00 \\
        \hline
         1a & 1 & Short & 1.55 & 2.02 & 95.46 \\
        1a & 2 & Short & 1.42 & 1.99 & 95.81 \\
        1a & 2 & Long & 2.00 & 2.80 & 94.92 \\
        1a & 3 & Short & 0.94 & 1.35 & 95.44 \\
        \hline
         1b & 1 & Short & 1.39 & 1.92 & 94.83 \\
        1b & 2 & Short & 1.14 & 2.28 & 94.94 \\
        1b & 2 & Long & 2.20 & 4.15 & 94.75 \\
        1b & 3 & Short & 0.69 & 1.35 & 94.98 \\
        \hline
         1c & 1 & Short & 2.78 & 3.07 & 95.60 \\
        1c & 2 & Short & 3.98 & 3.78 & 90.97 \\
        1c & 2 & Long & 5.92 & 5.13 & 85.37 \\
        1c & 3 & Short & 2.44 & 2.29 & 91.71 \\
        \hline
         2a & 1 & Short & 3.61 & 4.40 & 94.13 \\
        2a & 2 & Short & 3.68 & 4.71 & 93.25 \\
        2a & 2 & Long & 5.48 & 6.88 & 89.81 \\
        2a & 3 & Short & 3.53 & 3.54 & 89.95 \\
        \hline
         2b & 1 & Short & 10.43 & 1.57 & 63.99 \\
        2b & 2 & Short & 12.18 & 1.57 & 62.02 \\
        2b & 2 & Long & 13.15 & 1.72 & 58.38 \\
        2b & 3 & Short & 12.18 & 1.66 & 55.32 \\
        \hline
         3a & 1 & Short & 10.25 & 6.19 & 73.93 \\
        3a & 2 & Short & 10.60 & 5.80 & 70.80 \\
        3a & 2 & Long & 11.92 & 9.74 & 67.42 \\
        3a & 3 & Short & 8.96 & 5.34 & 68.83 \\
        \hline
        3b & 1 & Short & 3.66 & 4.62 & 94.15 \\
        3b & 2 & Short & 3.55 & 4.70 & 93.91 \\
        3b & 2 & Long & 4.97 & 6.99 & 91.38 \\
        3b & 3 & Short & 2.98 & 3.56 & 92.23 \\
        \hline
        3c & 1 & Short & 212.89 & 69.43 & 0.65 \\
        \hline
		\hline
	\end{tabular}
	\label{tab:supp_rcomp}
\end{table}

\begin{table}[H]
	\caption{Performance metrics for identification of additional parameters in Scenario 2a.}
	\centering
	\begin{tabular}{l|lll}
		\hline
		Parameter & MSE ($10^{-2}$) & Bias ($\%$) & Coverage Probability ($\%$) \\
		\hline
        $\beta_0$ & 2.42  & -1.39  & 96.05 \\
        $\beta_1$ & 5.11  & -2.82  & 95.11 \\
        $\boldsymbol{\theta}$ & 0.83  & 1.01  & 89.22 \\
        $\boldsymbol{\lambda}$ & 9.76   & 30.54  & 94.93 \\
        $\mathbf{w^*}$ & 1.70   & 3.28  & -$^*$ \\
		\hline
	\end{tabular}
 
        {\footnotesize $^*$Credible intervals are generated for $\boldsymbol{\lambda}$ rather than $\mathbf{w^*_s}$}
	\label{tab:paramident}
\end{table}

\end{document}